\documentclass[
 reprint,
 superscriptaddress,
 amsmath,amssymb,
 aps,
 prb,
]{revtex4-1}

\usepackage{amsmath,leftidx}
\usepackage{bm}
\usepackage{bbm}
\usepackage{times}
\usepackage{CJK}
\usepackage{color}
\usepackage[normalem]{ulem}
\usepackage{physics}
\usepackage{braket}

\makeatletter
\let\Caption\@makecaption
\makeatother
\usepackage{subcaption}
\makeatletter
\let\@makecaption\Caption
\makeatother

\usepackage{graphicx}

\begin{document}


\begin{CJK*}{UTF8}{}
\title{Impurity effects in one dimensional spin nematic liquid}
 \CJKfamily{bsmi}

\author{Fumiya Nakamura} 
\affiliation{Quantum Matter Program, Graduate School of Advanced Science and Engineering, Hiroshima University,
Higashihiroshima, Hiroshima 739-8530, Japan}

\author{Yasuhiro Tada}
\email[]{ytada@hiroshima-u.ac.jp}
\affiliation{Quantum Matter Program, Graduate School of Advanced Science and Engineering, Hiroshima University,
Higashihiroshima, Hiroshima 739-8530, Japan}
\affiliation{Institute for Solid State Physics, University of Tokyo, Kashiwa 277-8581, Japan}


\begin{abstract}
We study impurity effects in the spin nematic phase of 
the $S=1/2$ $J_1$-$J_2$ frustrated spin chain under an external magnetic field
by using the infinite density matrix renormalization group and the bosonization. 
It is found that local magnetization almost saturates around the impurity and
the entanglement entropy nearly vanishes at the corresponding bonds, not only when the magnetic interactions near 
the impurity are weakened but also when they are strengthened compared to those in the bulk.
Then, we examine spin correlations and Friedel oscillations induced by the impurity.
The bosonization provides a qualitative understanding of the characteristic behaviors of the magnetization.
We also discuss impacts of the impurity on experiments by focusing on NMR spectra. 
\end{abstract}


\maketitle
\end{CJK*}

\section{Introduction}
\label{introduction}

The spin nematic state is characterized by a vanishing spontaneous dipole moment and 
a non-zero spin quadrupole moment, where the time reversal symmetry is preserved but the rotational symmetry in spin space is broken. Under a magnetic field in the $z$-direction which explicitly breaks the time reversal symmetry, the spin nematic state is characterized by $\braket{S^+}=0$ and $\braket{S^+S^+}\neq 0$. The quadrupole moment is defined by two spins at neighboring sites in an $S=1/2$ system and correspondingly 
the spin nematic phase is characterized as a bond order.

Theoretically, the spin nematic states under magnetic fields are understood as a Bose-Einstein condensation 
of two-magnon bound pairs~\cite{heidrich2006frustrated, kecke2007multimagnon, hikihara2008vector, zhitomirsky2010magnon, shannon2006nematic}. Indeed, numerical calculations have confirmed that the magnetization has a magnetic field dependence that varies in steps of $\Delta S^z = 2$. 
The operators that create (annihilates) a magnon pair can be described by the spin quadrupole order $Q^{\alpha \beta}$ as $S^-S^-=Q^{x^2-y^2}-iQ^{xy}$ ($S^+S^+=Q^{x^2-y^2}+iQ^{xy}$), 
which bridges between the two-magnon bound pairs and the quadrupole moments. A model for the spin nematic phase under a magnetic field is the $S=1/2$ $J_1$-$J_2$ frustrated model with the ferromagnetic (FM) nearest-neighbor and antiferromagnetic (AFM) next nearest-neighbor interactions~\cite{chubukov1991chiral, dmitriev2006frustrated, heidrich2006frustrated, kecke2007multimagnon, vekua2007correlation, hikihara2008vector, sudan2009emergent, sato2013spin}. There are several 
quasi one-dimensional materials that are described by this model, such as LiCuV$\text{O}_4$~\cite{enderle2005quantum, orlova2017, svistov2011new, mourigal2012evidence, buttgen2012high, masuda2011spin, nawa2013anisotropic}, LiCuSb$\text{O}_4$~\cite{dutton2012quantum}, PbCuS$\text{O}_4\text{(OH)}_2$~\cite{yasui2011multiferroic, wolter2012magnetic}, $\text{Rb}_2\text{Cu}_2\text{Mo}_3\text{O}_{12}$~\cite{hase2004magnetic}, and Li$\text{Cu}_2\text{O}_2$~\cite{masuda2005spin}
where spin nematic states are expected to be realized. 
Intuitively, the nearest-neighbor FM term $J_1$ contributes to the formation of magnon bound pairs, 
and the magnon pairs can move 
by the next-neighbor AFM term $J_2$. 
More precisely, the spin nematic state in the one-dimensional $J_1$-$J_2$ model can be understood as 
a Tomonaga-Luttinger liquid (TLL) of magnon pairs which is stabilized in a wide range of the model parameters,
$J_1/J_2 \gtrsim -2.7$, near the saturation field. 
In an intermediate field region, the spin-density-wave (SDW) is stable and these two states are characterized 
by quasi long range orders of corresponding spin correlations.

Although there have been several theoretical proposals for an experimental observation of a spin nematic state~\cite{sato2009nmr, sato2011field, smerald2016theory, onishi2015magnetic, michaud2011theory, furuya2017angular, furuya2018electron}, 
it is very difficult to obtain direct experimental evidence. 
This is basically because a spin nematic state is not characterized by local magnetization but by 
a quadrupole moment $Q^{\alpha\beta}$.
The former can be observed by standard experimental techniques, but there is no experimental probe which directly couples 
with a quadrupole moment.
In spite of these difficulties, 
there have been acccumulated experimental evidence for the existence of spin nematic states in high field regions.
For example, in the NMR experiment on LiCuVO$_4$, 
there exists a phase below the saturation field 
where local magnetization is uniform and its value depends on the magnetic field~\cite{orlova2017}, 
which is distinguished from both an SDW state and a fully polarized state.
On the other hand, such a behavior was not clearly observed in the earlier NMR experiments and it was argued that
there were subtle impurity effects which caused some difficulties in the interpretation 
of the experimental results~\cite{buttgen2014search}.

In general, local impurity effects have been used as probes to analyze low energy properties in 
both gapped~\cite{hagiwara1990,tedoldi1999,yoshida2005} and gapless spin systems~\cite{eggert1995,takigawa1997}. 
Theoretical studies of impurity effects in TLL have been perfomed by Kane and Fisher and others with the renormalization group analysis~\cite{kane1992transport, kane1992resonant, kane1992transmission, furusaki1993resonant, furusaki1993single}. 
In particular, for a TLL with a single barrier, 
the physical consequence crucially depends on the TLL parameter $K $ (Kane-Fisher problem):
for $K < 1$, the system is effectively decoupled into two chains, while for $K > 1$, 
the impurity becomes negligible at low energies. 
Detailed behaviors of local magnetization and the Friedel oscillations have also been studied by 
field theories~\cite{eggert1995, egger1995friedel, egger1996friedel}.
Such characteristic responses to impurties can be used for probing gapless spin systems as demonstrated 
in the NMR experiment~\cite{takigawa1997}.
However, there are only a few studies for impurity effects in the spin nematic states~\cite{takano2011} 
and further theoretical studies are desirable.

In this study, we discuss impurity effects in the one dimensional $J_1$-$J_2$ model by using 
the infinite density matrix renormalization group (iDMRG)~\cite{schollwock2011density, mcculloch2007density, tenpy, lo2019crossover, phien2012infinite} 
and the phenomenological bosonization~\cite{kecke2007multimagnon, hikihara2008vector},
where the impurity is modeled by local bond disorder. 
We show that local magnetization almost saturates around the impurity and
the entanglement entropy nearly vanishes at the corresponding bonds, not only when the magnetic interactions near 
the impurity are weakened but also when they are strengthened compared to those in the bulk.
We then examine spin correlations and Friedel oscillations, and provide a field theoretical understanding of 
the numerical results.
Implications for NMR experiments are also discussed.

This paper is organized as follows. In Sec.\ref{sec:Model}, we introduce the model with 
an impurity. In Sec.\ref{sec:Result}, we first examine the local magnetization and the entanglement entropy
around the impurity.
The spin correlations and Friedel oscillations are also investigated.
Then we provide a field theoretical understanding of our numerical results based on the bosonization. 
Finally, we discuss NMR spectra corresponding to the magnetization profiles.
A summary is given in Sec.\ref{sec:Summary}.

\section{Model and Numerical Method}
\label{sec:Model}

\begin{figure}[htb]
	\includegraphics[width=0.37\textwidth]{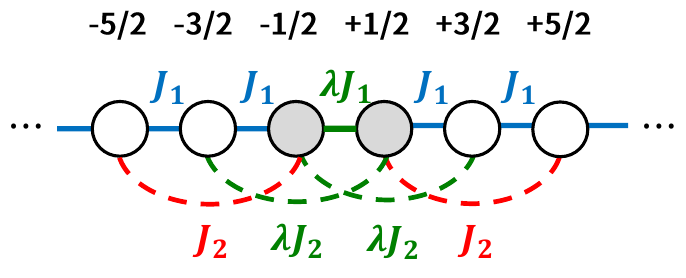}
	\caption{Schematic picture of the impurity system. The parameter $\lambda$ characterizes the magnitude of the exchange interaction around the sites $j=-1/2$ and $j=1/2$. 
}
	\label{set1}
\end{figure}

We consider an $S=1/2$ $J_1$-$J_2$ spin chain with infinite length, 
where impurities are described by a local change  
of interaction strength (i.e. bond disorder) as shown in Fig.\ref{set1}. 
The Hamiltonian is $\mathcal{H} = \mathcal{H}_{\text{bulk}} + \mathcal{H}_{\text{imp}}$,
\begin{align}
	\mathcal{H}_{\text{bulk}} &=  J_1 \sum_j \bm{S}_j \cdot \bm{S}_{j+1} + J_2 \sum_j \bm{S}_j \cdot \bm{S}_{j+2} - h_z \sum_j S^{z}_j,
	\label{bulk1}\\
    \mathcal{H}_{\text{imp}} &=  (\lambda-1) J_1 \bm{S}_{-\frac{1}{2}} \cdot \bm{S}_{\frac{1}{2}} \nonumber \\
    &+ (\lambda-1) J_2 \left( \bm{S}_{-\frac{3}{2}} \cdot \bm{S}_{\frac{1}{2}} + \bm{S}_{-\frac{1}{2}} \cdot \bm{S}_{\frac{3}{2}} \right),
\label{eqimp1}
\end{align}
\noindent
where $S_j^{\mu}$ is the spin operator at site $j\in {\mathbb Z}+1/2$. 
$J_1<0, J_2>0$ are FM nearest neighbor interaction and AFM next nearest neighbor interaction, 
and $h_z$ is an external magnetic field. 
In this Hamiltonian, the impurities are supposed to be located around $j\sim 0$ and are
simply modelded by the additional interaction ${\mathcal H}_{\rm imp}$ 
characterized by the parameter $\lambda$, where $\lambda=1$ corresponds to a uniform chain without any impurity. 
For $\lambda<1$, the exchange interaction near the impurity is weakened, 
while for $\lambda>1$ it is strengthened. 
In this study, the bond $\{-1/2,1/2\}$ is called the impurity bond. 
This impurity model allows one to discuss impurity effects in a coherent way by changing the 
pamameter $\lambda$. 
Note that there is a single impurity bond in the present infinite chain system.
We also consider a different impurity model in Appendix \ref{app:defect} which corresponds to
the previous NMR study \cite{buttgen2014search}. 
In this study, we consider two representative values of the parameter $J_1/J_2$,
namely $J_1/J_2=-0.5$ and $J_1/J_2=-2.0$ in Eq.~\eqref{eqimp1}.
It turns out that calculation results for these two parameters are qualitatively same, and therefore
we focus on $J_1/J_2=-0.5$ and briefly touch on $J_1/J_2=-2.0$.
For the uniform chain with $\lambda=1$,
the system exhibits the quasi long range order of the spin nematic state at the average magnetization $M\gtrsim0.35$  
and that of the SDW  state at $M\lesssim0.35$ according to the previous studies \cite{hikihara2008vector, sudan2009emergent},
which we have confirmed in our uniform infinite chain.

In this work, we perform the iDMRG calculations developed in the previous studies
with using the TeNPy library~\cite{tenpy, phien2012infinite, lo2019crossover}. 
In this method, an infinite inhomongeneous chain is divided into three regions, namely, the left semi-infinite chain, 
right semi-infinite chain, and  ``window" region with a finite length $L$ in the middle.
The left and right chains behave as semi-infinite environments for the window chain.
It is stressed that there is no open ends in an infinite chain and finite size effects are greatly suppressed in this method.
As a result, it enables accurate and efficient calculations of an impurity system.
In the numerical calculations, the average magnetization per site 
in each of the three regions is fixed to be common values $M=0.2$ and $M=0.4$ corresponding to the SDW phase
and spin nematic phase, respectively.
We use the bond dimension $\chi=400$, for which
the truncation error is $10^{-9}\sim10^{-11}$ in the spin nematic phase and 
is $\sim 10^{-8}$ in the SDW phase. 
Before doing the calculations in the window, we perform iDMRG for a uniform system to obtain environment tensors 
corresponding to the left and right semi-infinite chains. 
These environment tensors are fixed as the boundaries of the window chain.
We find that there are some sutble effects of the finite bond dimension $\chi$ and the finite window size $L$,
but main conclusions of our study are essentially unchanged for lager $\chi$ and $L$ (see Appendix~\ref{app:finite_size}).


\section{Results}
\label{sec:Result}
In this section, we show results of the numerical calculations for the window region with $L=400$
connected to the left and right semi-infinite chains.
We first discuss profiles of magnetization and entanglement entropy to understand local configurations
around the impurity bond $\{ -1/2,1/2\}$.
Then, we examine spin correlations between neighboring sites and Friedel oscillations induced by the impurity.
We provide a field theoretical picture for the numerical results.
Finally, experimental implications for NMR spectra are discussed.


\subsection{Magnetization and entanglement entropy}
\label{sec:Magnetization}
\begin{figure}[htb]
    \begin{subfigure}{0.23\textwidth}
        \includegraphics[width=\linewidth]{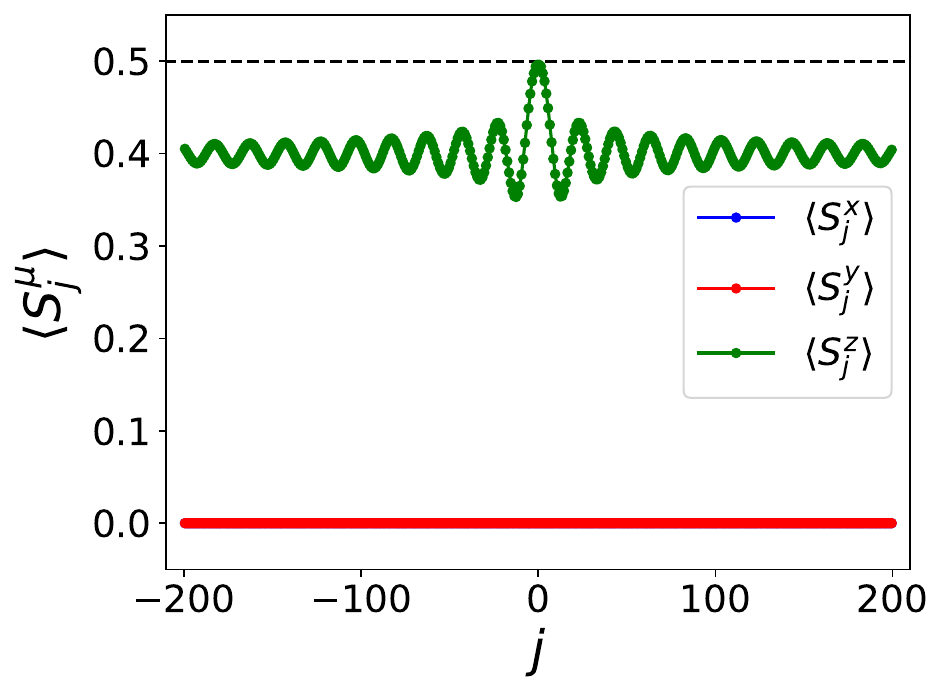}
        \caption{$\lambda=0.8$}
        \label{mag1}
    \end{subfigure}
    \begin{subfigure}{0.23\textwidth}
        \includegraphics[width=\linewidth]{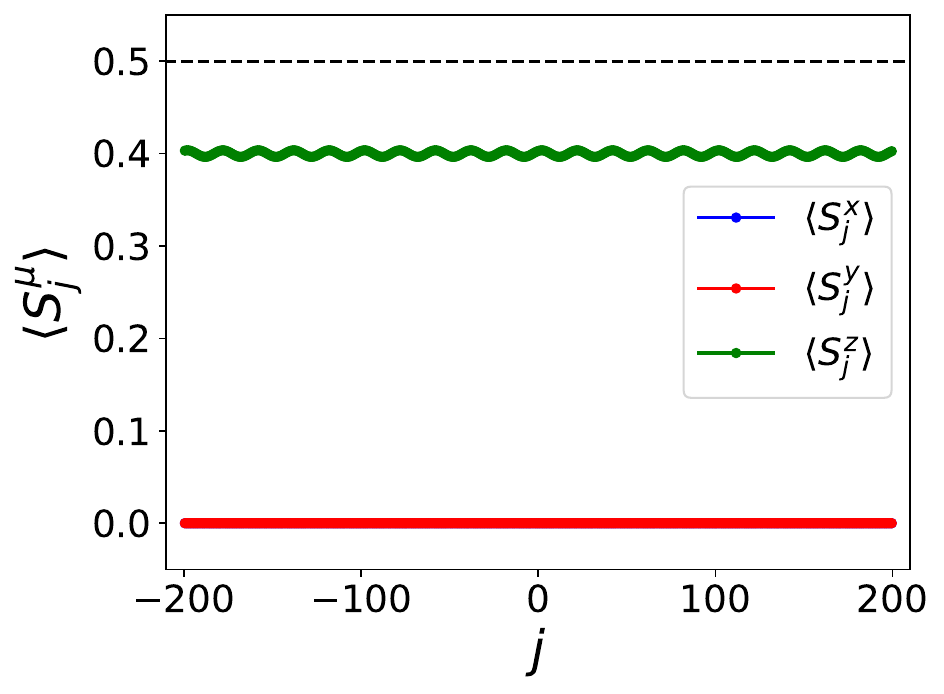}
        \caption{$\lambda=1.0$}
        \label{mag2}
    \end{subfigure}
    \begin{subfigure}{0.23\textwidth}
        \includegraphics[width=\linewidth]{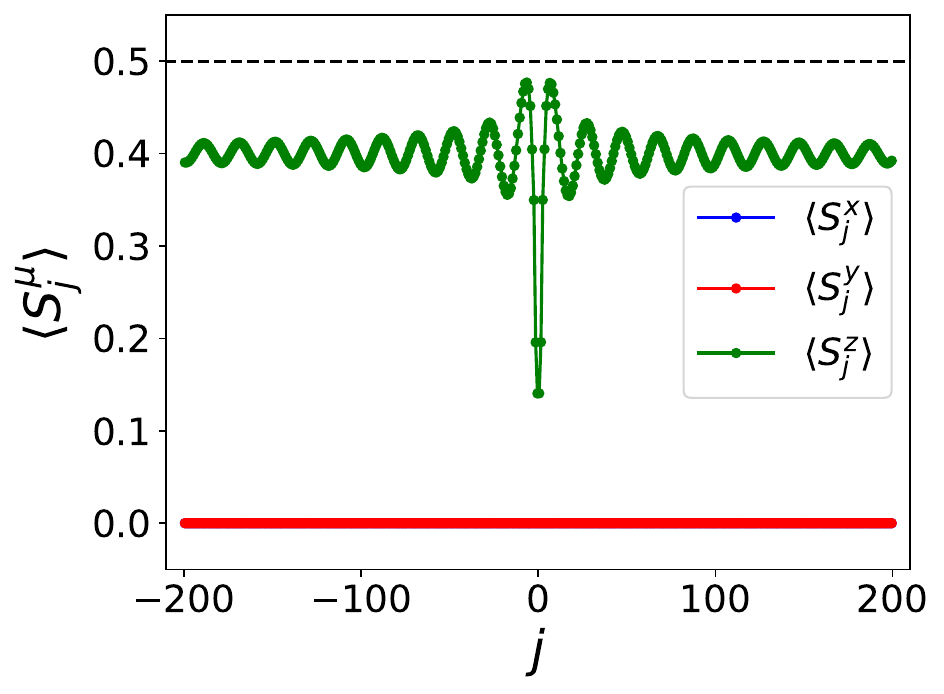}
        \caption{$\lambda=1.2$}
        \label{mag3}
    \end{subfigure}
    \begin{subfigure}{0.23\textwidth}
        \includegraphics[width=\linewidth]{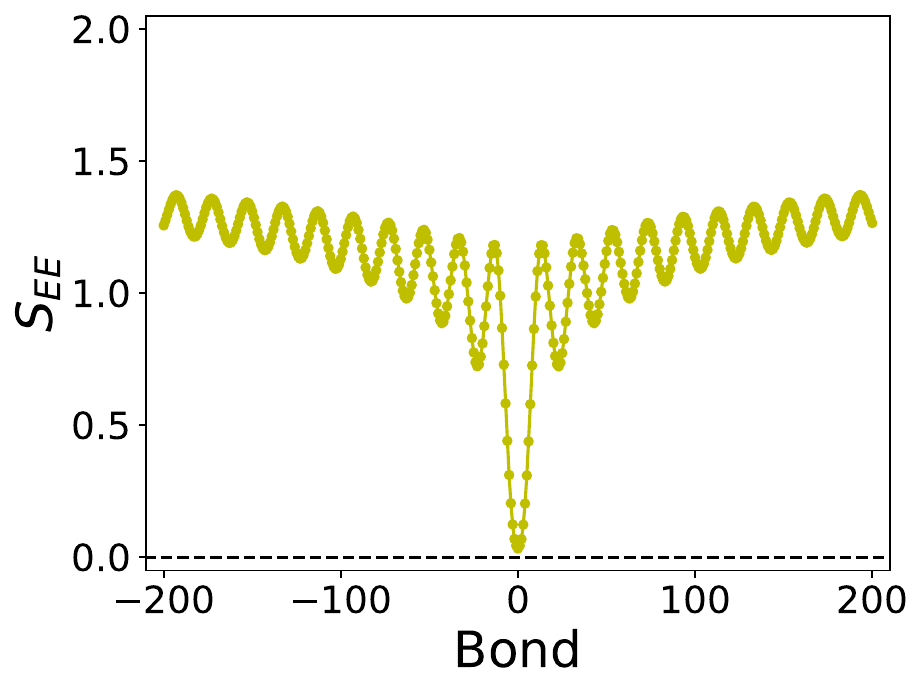}
        \caption{$\lambda=0.8$}
        \label{en1}
    \end{subfigure}
    \begin{subfigure}{0.23\textwidth}
        \includegraphics[width=\linewidth]{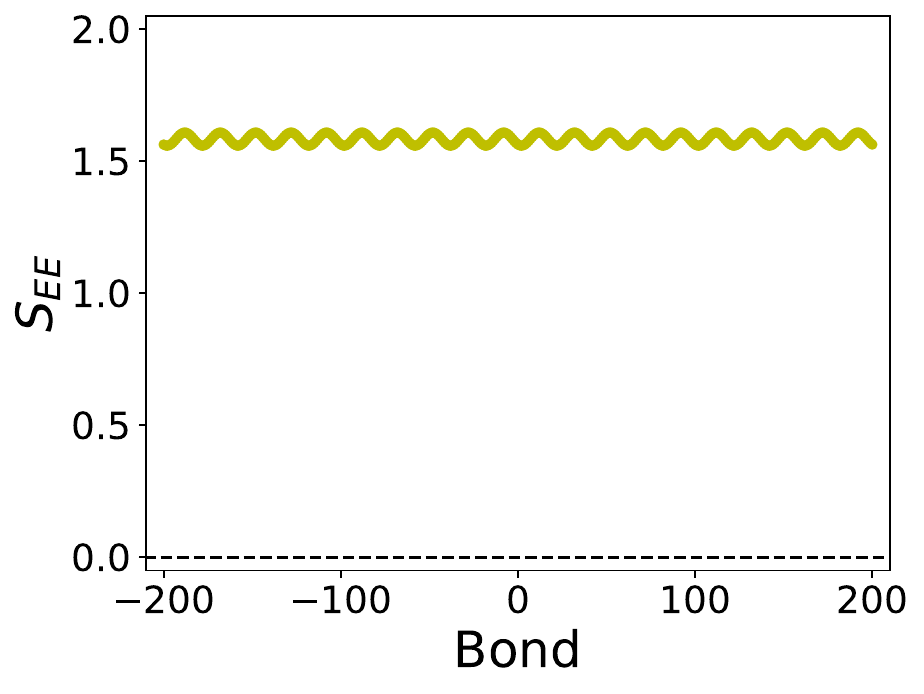}
        \caption{$\lambda=1.0$}
        \label{en2}
    \end{subfigure}
    \begin{subfigure}{0.23\textwidth}
        \includegraphics[width=\linewidth]{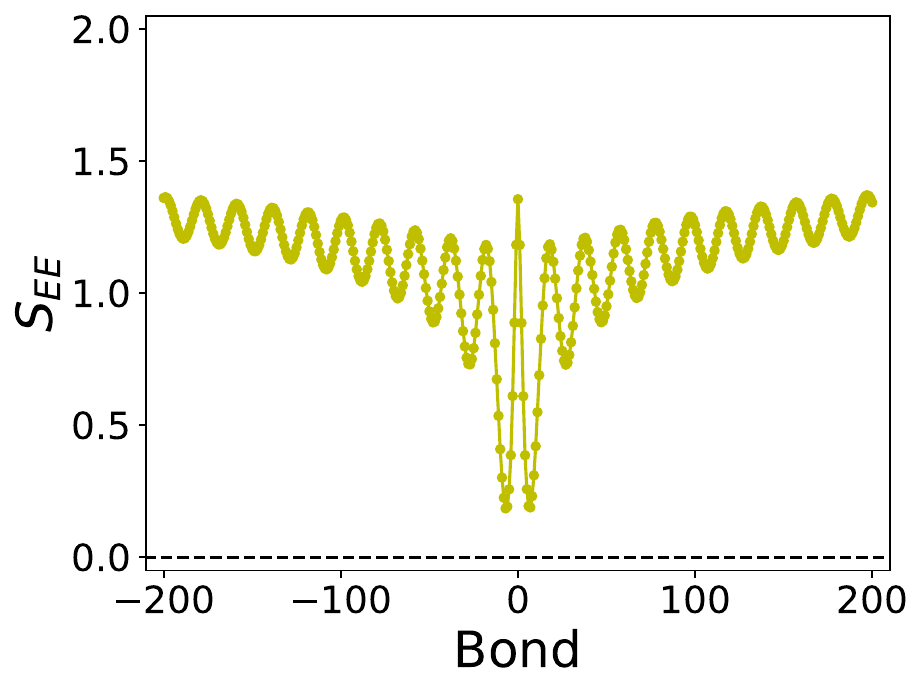}
        \caption{$\lambda=1.2$}
        \label{en3}
    \end{subfigure}
    
    \caption{(a) (b) (c) The local spin polarization $\braket{S^\mu_j} (\mu = x, y, z)$ and (d) (e) (f) the entanglement entropy $S_{\text{EE}}$ for $L = 400$ with $J_1/J_2=-0.5$ and $M=0.4$. 
The dashed lines indicate $M = 0.5$ in (a), (b), (c), and $S_{\text{EE}} = 0.0$ in (d), (e), (f), respectively. 
Results for $\lambda=0.8, 1.0, 1.2$ are shown.}
    \label{magenall}
\end{figure}
We discuss the local magnetic moment $\braket{S_j^{\mu}}$ at each site $j$ and the entanglement entropy
between sites $j$ and $j+1$,
\begin{align}
S_{EE}(j)=-{\rm Tr}_{A_j}\rho_{A_j}\log \rho_{A_j},
\end{align}
where $\rho_{A_j}$ is the reduced density matrix for the subsystem $A_j=\{\cdots, j-2,j-1,j\}$.
The argument $j$ corresponds to the bond $\{j,j+1\}$ and 
we often use the same index $j$ to represent both sites and nearby bonds for brevity. 
$S_{EE}(j)$ is a constant in a uniform ground state for given parameters, 
while it depends on the position $j$ in a non-uniform state.

Figure \ref{magenall} shows the local magnetization and the entanglement entropy for the window region 
in the spin nematic phase at $M=0.4$. 
The local spin polarization $\braket{S^{x, y}_j}$ is zero at any sites, 
while $\braket{S_j^z}$ exhibits spatial oscillations around its average $M=0.4$.
The weak oscillations of $\braket{S_j^z}$ for the uniform coupling case ($\lambda=1$) are numerical artifacts 
mainly due to the finite bond dimension $\chi$, 
and we have confirmed that they become smaller as the bond dimension $\chi$ is increased 
(see Appendix~\ref{app:finite_size}).
Another numerical artifact is that $\braket{S_j^z}$ and $S_{EE}(j)$ in the window region are not smoothly connected 
to those in the semi-infinite chains at the sites $|j|> 200$.
It is found that 
this discontinuity does not affect physical behaviors in the window region which we are interested in 
(Appendix~\ref{app:finite_size}).
On the other hand, one can clearly see the Friedel oscillations around the impurity bond 
when $\lambda\neq1$, which is distinguished from the above artifacts.
Details of the Friedel oscillations will be discussed later in Sec.~\ref{sec:Friedel}.

For $\lambda < 1$, the magnetization near the impurity bond almost saturates, $\braket{S_j^z}\to 0.5$,
and correspondingly the entanglement entropy nearly vanishes $S_{EE}(j\sim0)\to0$.
On the other hand, for $\lambda > 1$, the magnetization very close to the impurity $\braket{S^z_{j\sim0}}$ 
is strongly suppressed
and the entanglement entropy remains moderate.
In this case, however, there are sharp oscillations around $j\sim0$ and the magnetization at sites slightly away from 
the impurity bond nearly saturates, $\braket{S_j^z}\simeq 0.5$, and the corresponding entanglement entropy
approaches zero.
The nearly satuated magnetization could be understood as a result of formation of spin-singlet-like pairs around the impurity as will be discussed in
Sec.~\ref{sec:Correlations}.
Although it may be rather natural that the spin chain is decoupled for the weakened bond with $\lambda<1$,
a similar decoupling takes place even for the strengthened bond with $\lambda>1$.
This implies that bond disorder would generally lead to similar effects of a spin chain in a coarse-grained scale.
In any case, we have the following properties for the present model,
\begin{equation}
	\braket{S^z_j} \rightarrow 0.5 \quad \mbox{and} \quad S_{EE}(j) \rightarrow 0
	\label{relationship1}.
\end{equation}
The relation between these two properties is not so trivial, but 
the ground state wavefunction $\ket{\Psi}$ must have a structure satisfying them.
If the local magnetization at sites $j$ and $j+1$ fully saturates, it has the structure 
on the bond $\{j,j+1\}$, 
\begin{align}
	\ket{\Psi}&=\sum_{\{\sigma_i\}_{i\neq j,j+1}} a({\{\sigma_i\}}_{i\leq j-1})b({\{\sigma_i\}}_{i\geq j+2}) \nonumber\\
&\qquad\qquad \times \ket{\cdots, \sigma_{j-1},\uparrow}\ket{\uparrow,\sigma_{j+2},\cdots},
\label{eq:psi}
\end{align}
where $\sigma_i=\uparrow,\downarrow$ is the spin index at site $i$
and $a({\{\sigma_i\}}_{i\leq j-1}),b({\{\sigma_i\}}_{i\geq j+2})$ are coefficients.
The entanglement entropy $S_{EE}(j)$ is clearly zero for this state, and the subsystems $A_{j}$ and its complement
are effectively decoupled.
The numerical results suggest that such effective decoupling takes place at the sites $j_{\rm dec}\simeq \pm 1/2$ for
$\lambda<1$ and $j_{\rm dec}\simeq\pm11/2$ for $\lambda>1$.

\begin{figure}[htb]

    \begin{subfigure}{0.23\textwidth}
        \includegraphics[width=\linewidth]{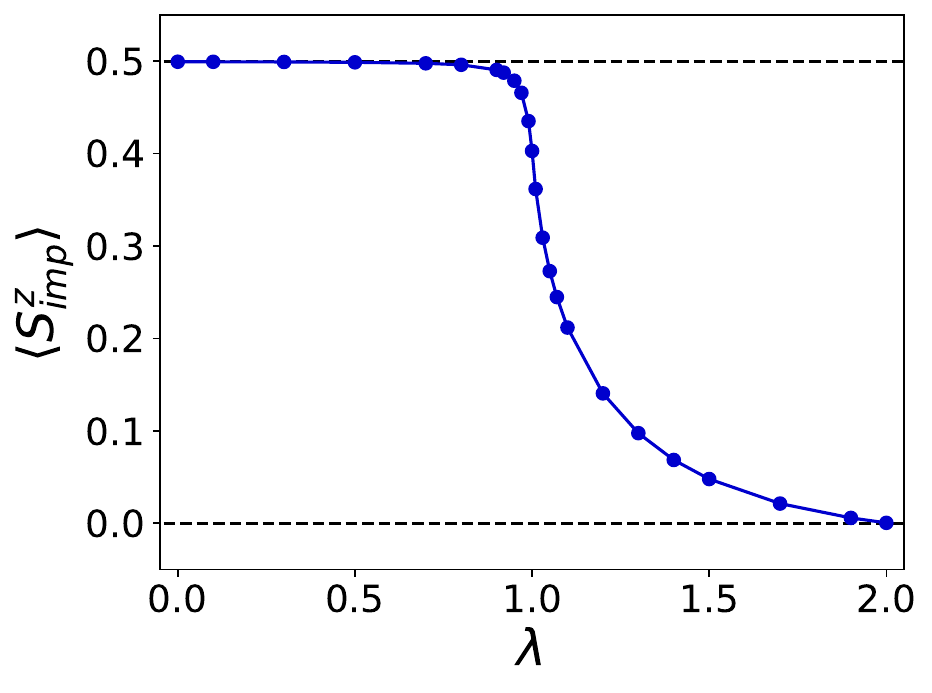}
        \caption{$j=1/2$}
        \label{maglam1}
    \end{subfigure}
    \begin{subfigure}{0.23\textwidth}
        \includegraphics[width=\linewidth]{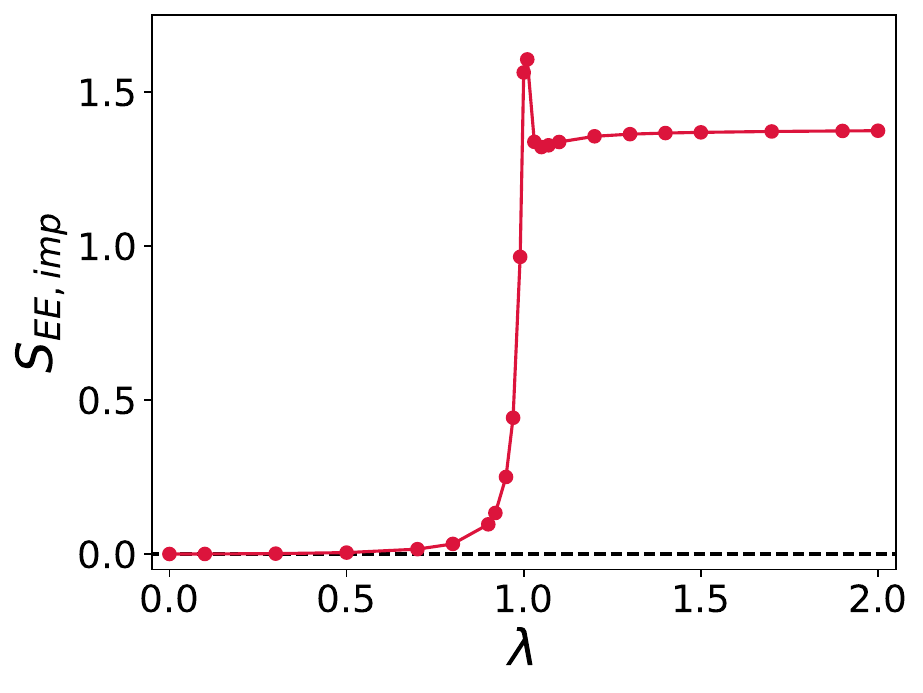}
        \caption{$j=1/2$}
        \label{enlam1}
    \end{subfigure}
    \begin{subfigure}{0.23\textwidth}
        \includegraphics[width=\linewidth]{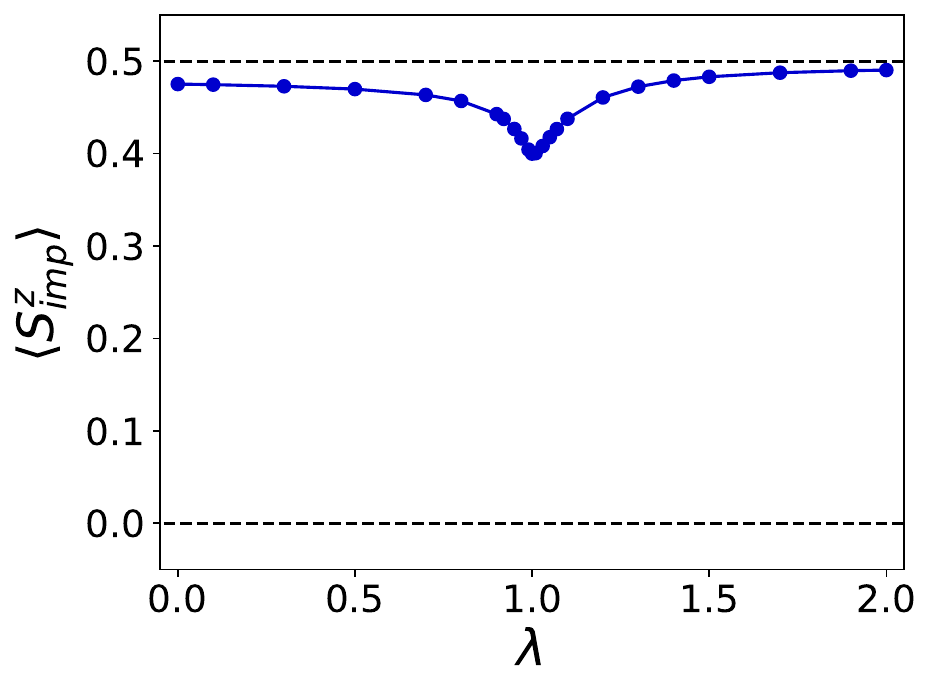}
        \caption{$j=11/2$}
        \label{maglam2}
    \end{subfigure}
    \begin{subfigure}{0.23\textwidth}
        \includegraphics[width=\linewidth]{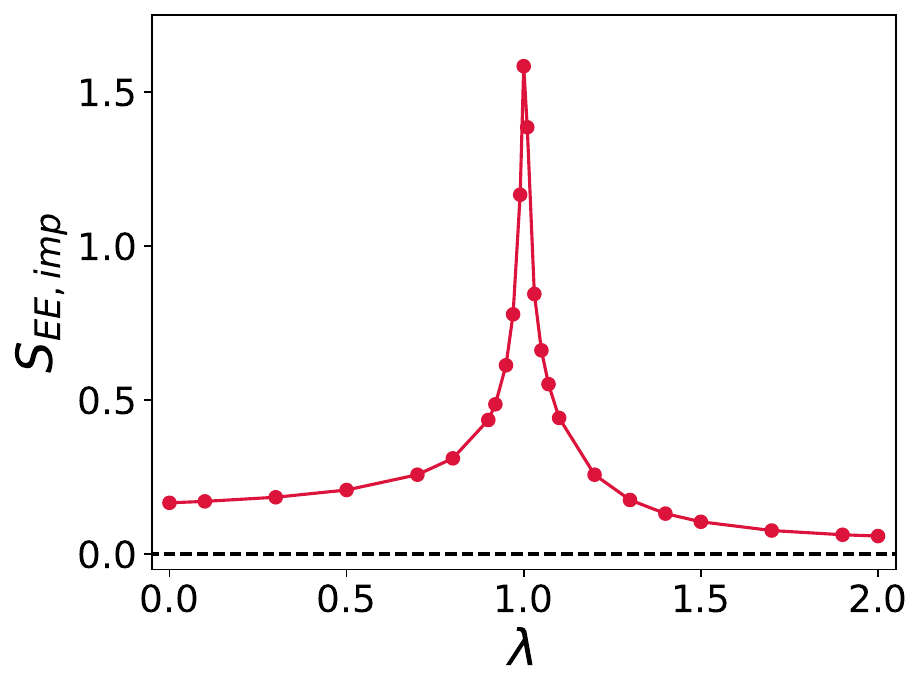}
        \caption{$j=11/2$}
        \label{enlam2}
    \end{subfigure}

    \caption{$\lambda$ dependence of the magnetization $\braket{S^z_{\text{imp}}} = (\braket{S^z_{-j}} + \braket{S^z_{j}})/2$
and entanglement entropy $S_{EE,\text{imp}}=(S_{EE}(-j)+S_{EE}(j))/2$ at the sites $\pm j$
near the impurity for $J_1/J_2=-0.5$ and $M=0.4$. 
(a), (b) The magnetization and the entanglement entropy for $j=1/2$.
The dashed lines indicate $M=0.5$ and $M=0$ for the magnetization and $S_{EE}=0$ for the entanglement entropy. 
(c), (d) The magnetization and the entanglement entropy for $j=11/2$.}
    \label{lamall}
\end{figure}

To further explore the relation between magnetization and entanglement entropy, 
we analyze their dependence on $\lambda$ in more detail. 
Figures \ref{maglam1} and \ref{enlam1} show the $\lambda$ dependence of magnetization and entanglement entropy near the impurity bond at the decoupling sites 
$j=\pm 1/2$. 
They are defined by $\braket{S^z_{\rm imp}}=(\braket{S^z_{-j_{\rm dec}}}+\braket{S^z_{j_{\rm dec}}})/2$ and 
$S_{EE,{\rm imp}}=(S_{EE}(j=-j_{\rm dec})+S_{EE}(j=j_{\rm dec}))/2$ with $j_{\rm dec}=1/2$.
In Fig.\ref{maglam1}, one can see that there is a sharp crossover of the magnetization $\braket{S^z_{\rm imp}}$
around $\lambda \simeq 1$. 
It rapidly increases and saturates for $\lambda<1$ and smoothly decreases down to nearly zero for $\lambda>1$.
Correspondingly, $S_{EE,{\rm imp}}$ vanishes for $\lambda<1$ and stay at some non-zero values for $\lambda>1$.
Similarly, Fig.\ref{maglam2} and Fig.\ref{enlam2} show the $\lambda$ dependence of magnetization and entanglement entropy at the other decoupling sites $j = \pm 11/2$.
In this case, the magnetization $\braket{S^z_{\rm imp}}$ with $j_{\rm dec} \simeq 11/2$ nearly saturates for $\lambda>1$
and the corresponding entanglement entropy $S_{EE,{\rm imp}}$ is strongly suppressed.
We have performed similar analyses for $J_1/J_2=-2.0$ and obtained qualitatively similar results with 
quantitatively stronger $\lambda$ dependence.
In any case,
one can numerically confirm the saturated magnetization and the vanishing entanglement entropy 
for a wide range of $\lambda$.
The characteristic $\lambda$-dependence of $\braket{S^z_{\rm imp}}$ around $\lambda\simeq1$ is
commonly seen in the Friedel oscillation as will be discussed in Sec.~\ref{sec:Friedel}.

\begin{figure}[htb]
    \begin{subfigure}{0.23\textwidth}
        \includegraphics[width=\linewidth]{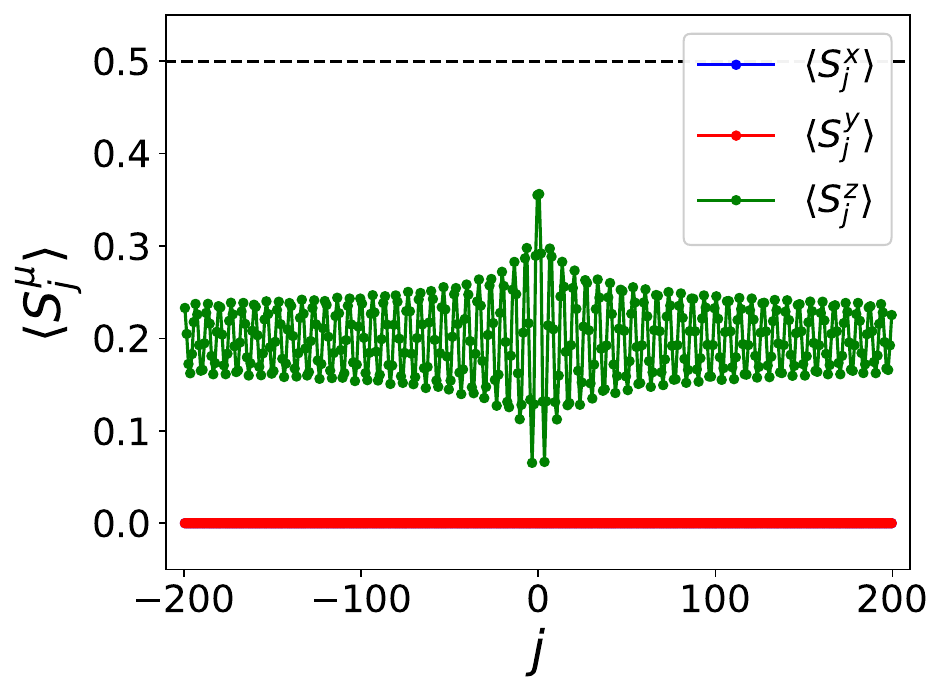}
        \caption{$\lambda=0.8$}
        \label{SDWmag1}
    \end{subfigure}
    \begin{subfigure}{0.23\textwidth}
        \includegraphics[width=\linewidth]{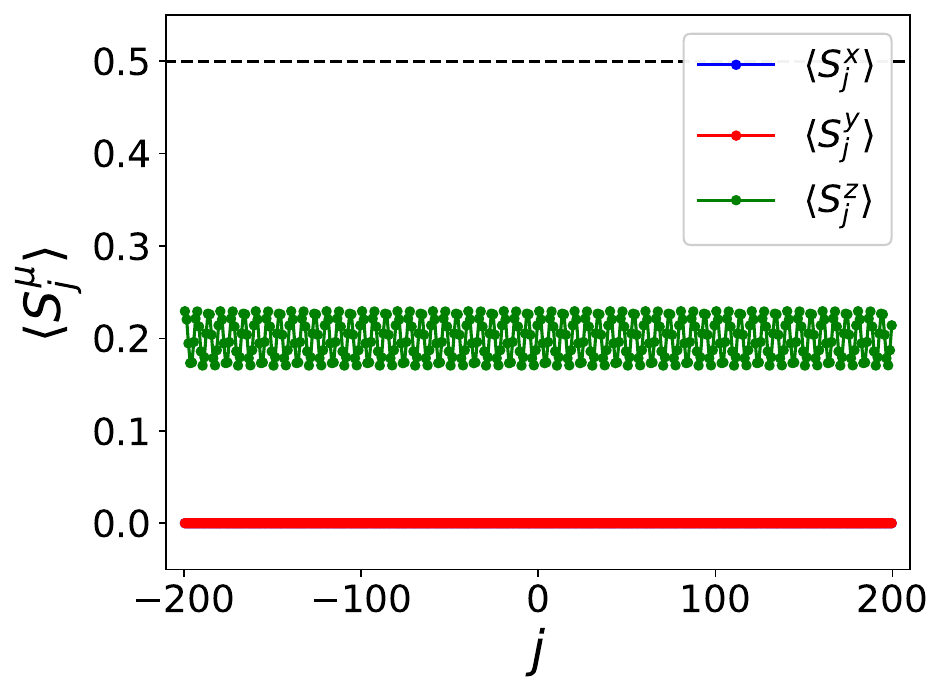}
        \caption{$\lambda=1.0$}
        \label{SDWmag2}
    \end{subfigure}
    \begin{subfigure}{0.23\textwidth}
        \includegraphics[width=\linewidth]{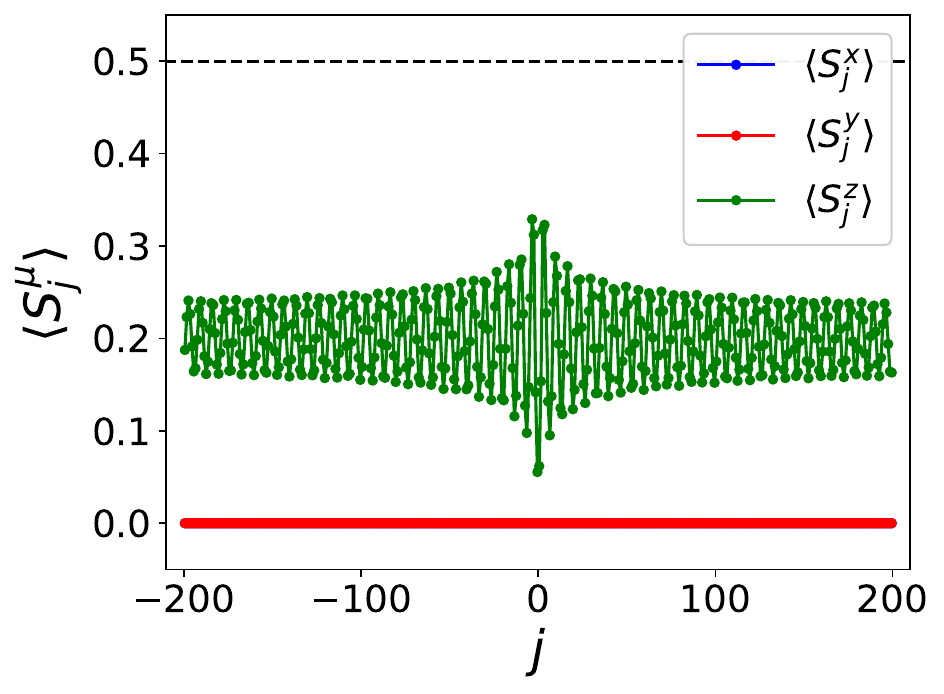}
        \caption{$\lambda=1.2$}
        \label{SDWmag3}
    \end{subfigure}
    \begin{subfigure}{0.23\textwidth}
        \includegraphics[width=\linewidth]{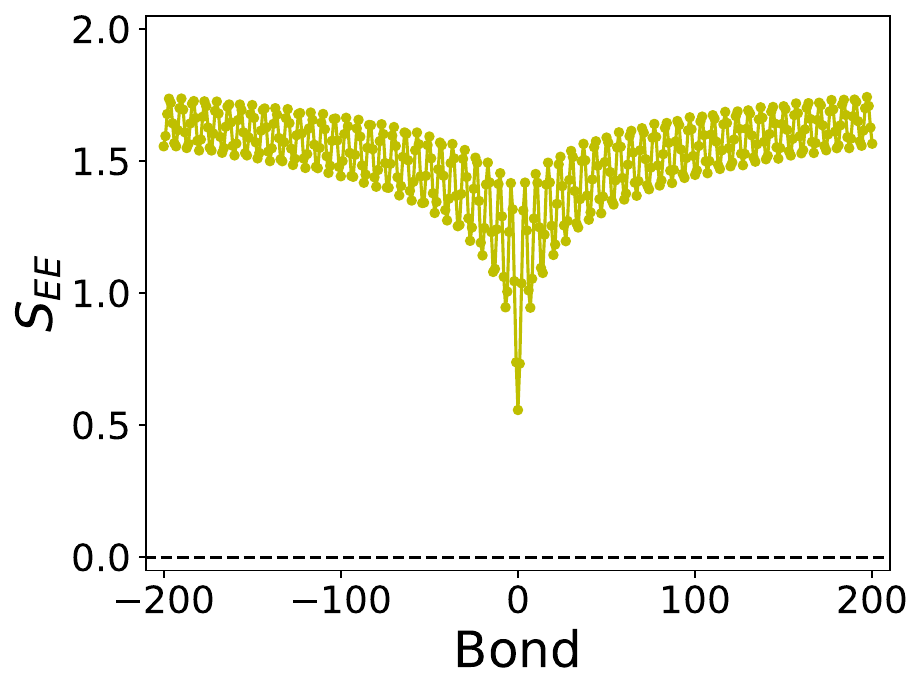}
        \caption{$\lambda=0.8$}
        \label{SDWen1}
    \end{subfigure}
    \begin{subfigure}{0.23\textwidth}
        \includegraphics[width=\linewidth]{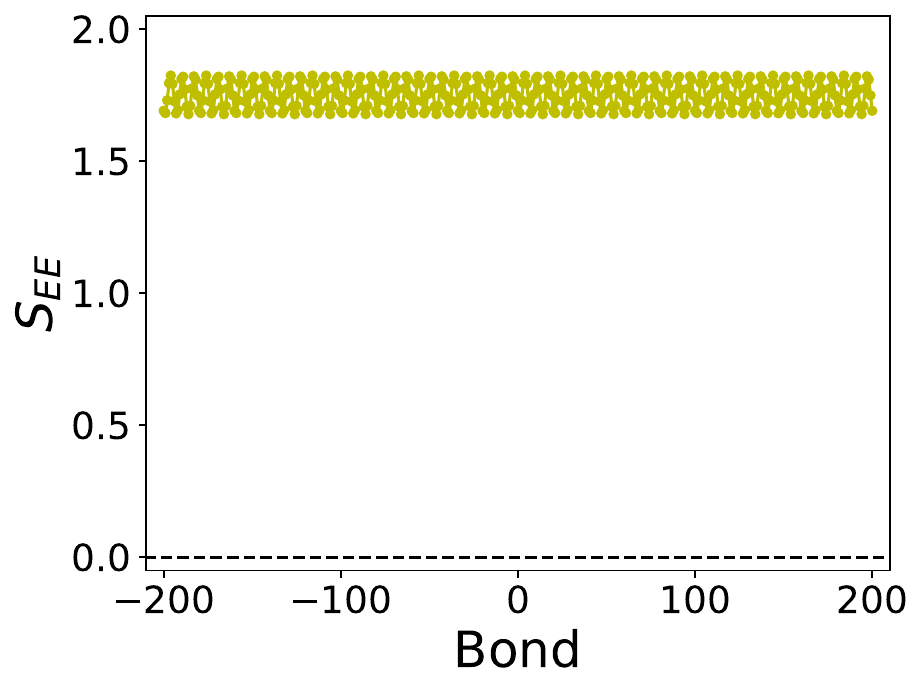}
        \caption{$\lambda=1.0$}
        \label{SDWen2}
    \end{subfigure}
    \begin{subfigure}{0.23\textwidth}
        \includegraphics[width=\linewidth]{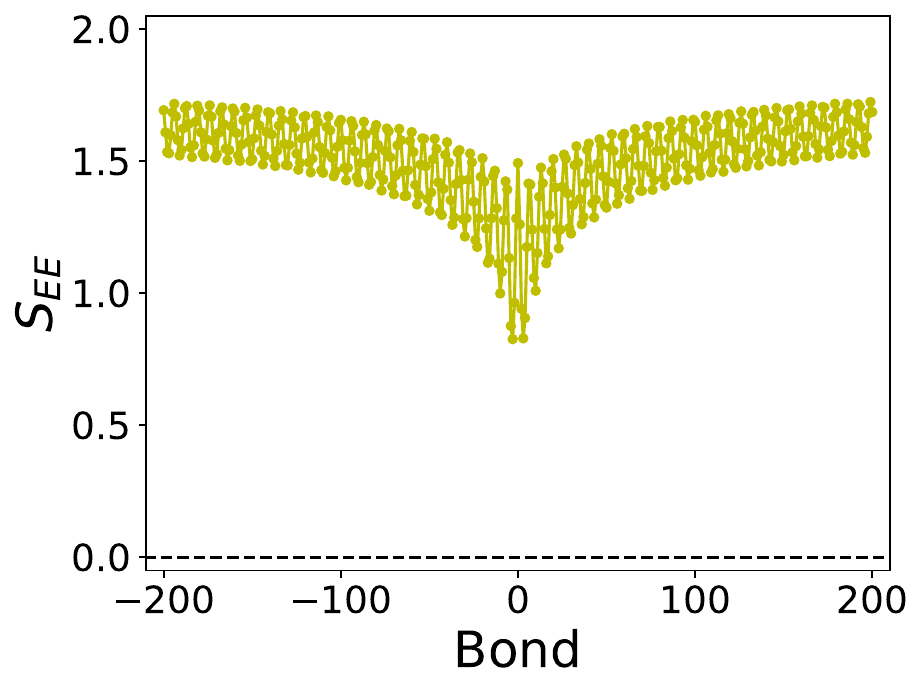}
        \caption{$\lambda=1.2$}
        \label{SDWen3}
    \end{subfigure}
    
    \caption{The local spin polarization $\braket{S^\mu_j} (\mu = x, y, z)$ and the entanglement entropy $S_{\text{EE}}$ for $L = 400$ with $J_1/J_2=-0.5$ and $M=0.2$ in the SDW phase. (a) (b) (c)  The local spin polarization 
and (d) (e) (f) the entanglement entropy for $\lambda=0.8, 1.0, 1.2$. }
    \label{SDWmagenall}
\end{figure}

We perform similar calculations for the SDW phase adjacent to the spin nematic phase. Figure \ref{SDWmagenall} 
shows the local magnetization and entanglement entropy of the impurity system in the SDW phase at $M=0.2$. 
Similarly to the spin nematic phase, 
some oscillations are seen even for the uniform Hamiltonian with $\lambda=1$ for
the SDW phase,
which is a finite bond dimension effect and is suppressed for large $\chi$ (Appendix \ref{app:finite_size}).
The oscillations of $\braket{S_j^z}$ around $\lambda = 1$ are simply amplified near the impurity bond for both $\lambda<1$ and $\lambda>1$ 
in a similar manner and there is no characteristic structure of the spatial profile
in contrast to the local magnetization in the spin nematic phase.
On the other hand, the entanglement entropy decreases when the corresponding magnetization is enhanced
as in the spin nematic phase.

Qualitative behaviors of magnetization for the spin nematic state and the SDW state  
share some common features, and there will be crosover between them in the present one-dimensional
system without a quantum phase transition. 
Besides, if the magnetization is very close to the saturation, $M\sim 0.5$, 
the impurity effects in the magnetization may be quantitatively suppressed, 
because the maximal value of the local magnetization is bounded by the saturation value $\braket{S_j^z}=0.5$.

For a comparison,
we have performed a similar analysis of magnetization also for the standard XXZ model (see Appendix \ref{app:XXZ}). 
The magnetization profile $\braket{S_j^z}$ in the XXZ model at $M=0.4$ 
is qualitatively similar to that in the spin nematic phase of the $J_1$-$J_2$ model, while $\braket{S_j^z}$
at $M=0.2$ is similar to that of the SDW phase.
This implies that local magnetization around the impurity is basically determined by the bulk magnetization $M$.
As will be discussed later, however,
there is some differences in the Friedel oscillations for the spin nematic state and the standard TLL,
which leads to quantitatively distinct behaviors in NMR spectra.

\subsection{Spin correlations}
\label{sec:Correlations}
\begin{figure*}[htb]

    \begin{subfigure}{0.33\textwidth}
        \includegraphics[width=\linewidth]{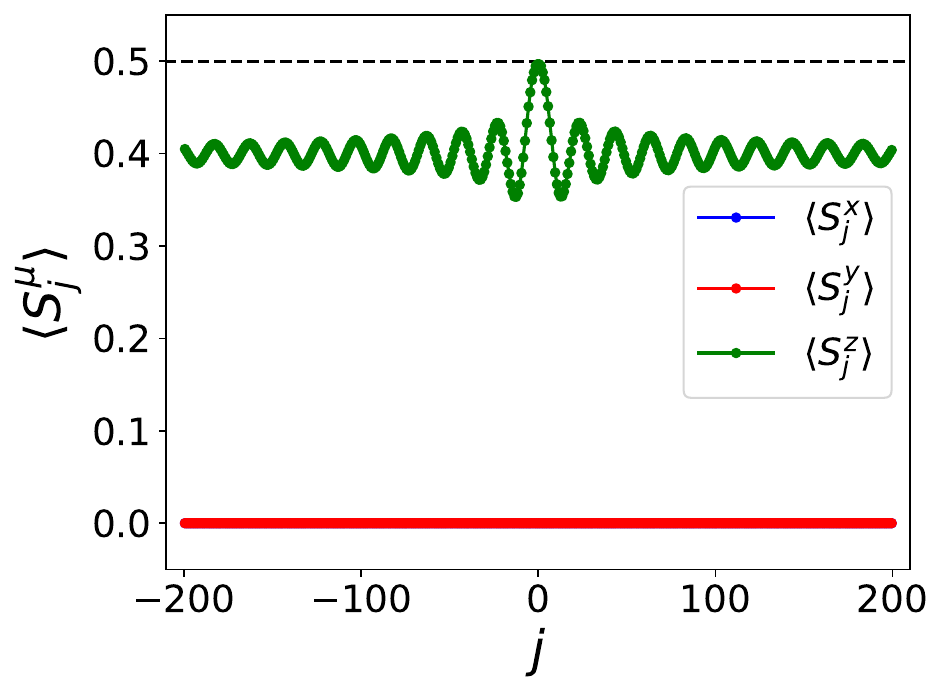}
        \caption{$\lambda_1=1.0$, $\lambda_2=0.8$}
        \label{FMmag}
    \end{subfigure}
    \begin{subfigure}{0.33\textwidth}
        \includegraphics[width=\linewidth]{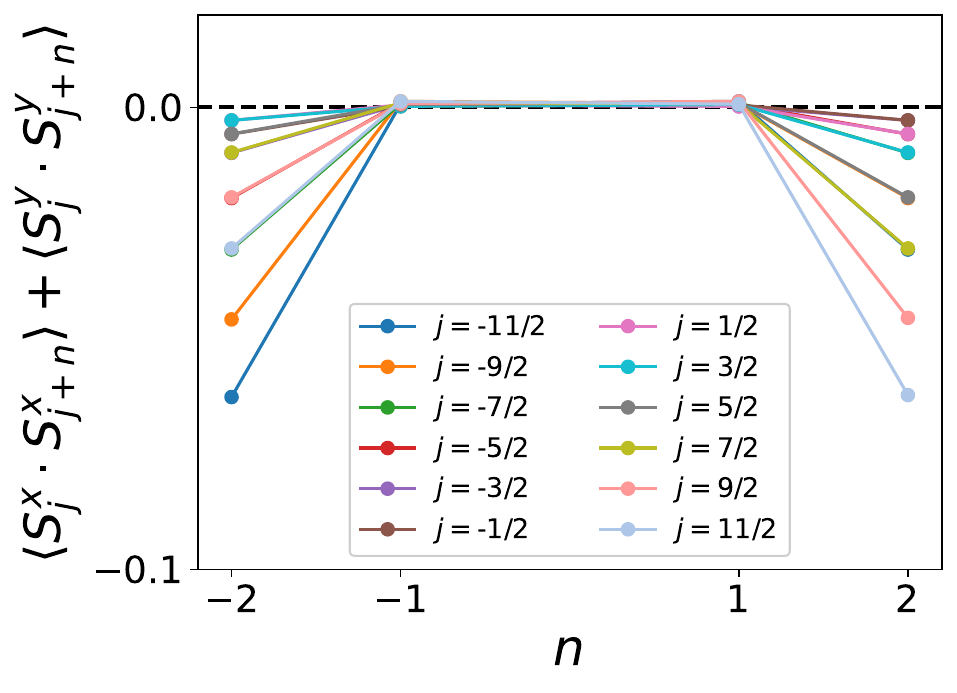}
        \caption{$\lambda_1=1.0$, $\lambda_2=0.8$}
        \label{FMCorXY}
    \end{subfigure}
    \begin{subfigure}{0.33\textwidth}
        \includegraphics[width=\linewidth]{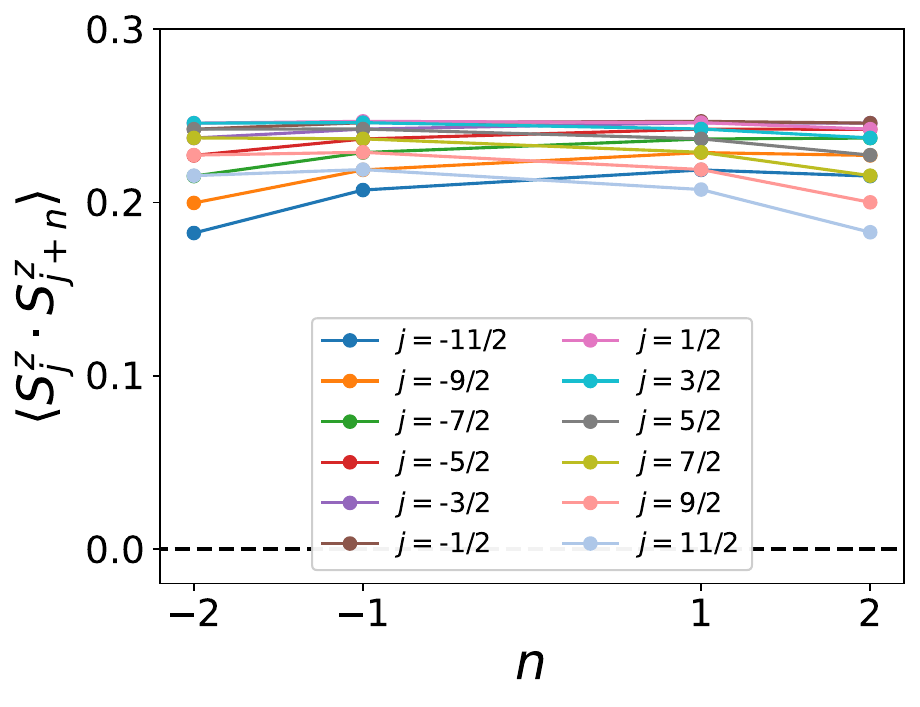}
        \caption{$\lambda_1=1.0$, $\lambda_2=0.8$}
        \label{FMCorZZ}
    \end{subfigure}
    \begin{subfigure}{0.33\textwidth}
        \includegraphics[width=\linewidth]{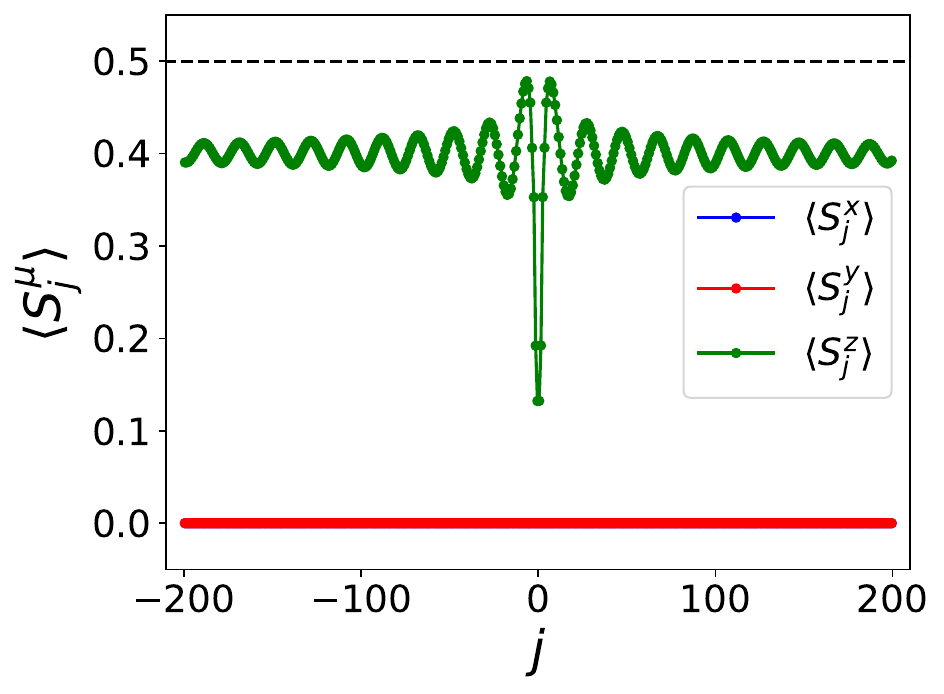}
        \caption{$\lambda_1=1.0$, $\lambda_2=1.2$}
        \label{AFMmag}
    \end{subfigure}
    \begin{subfigure}{0.33\textwidth}
        \includegraphics[width=\linewidth]{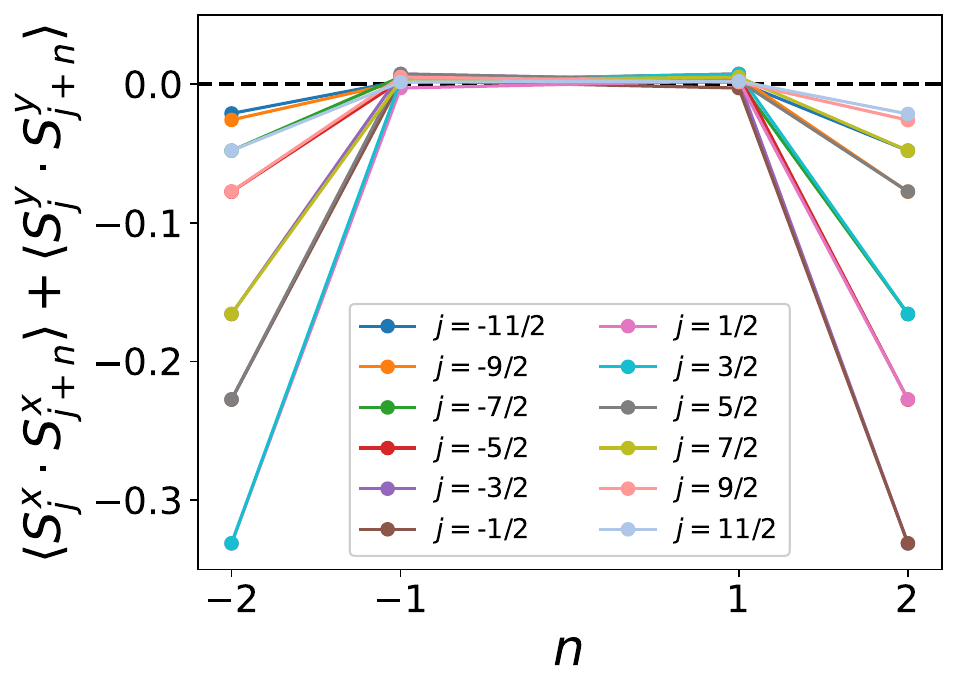}
        \caption{$\lambda_1=1.0$, $\lambda_2=1.2$}
        \label{AFMCorXY}
    \end{subfigure}
    \begin{subfigure}{0.33\textwidth}
        \includegraphics[width=\linewidth]{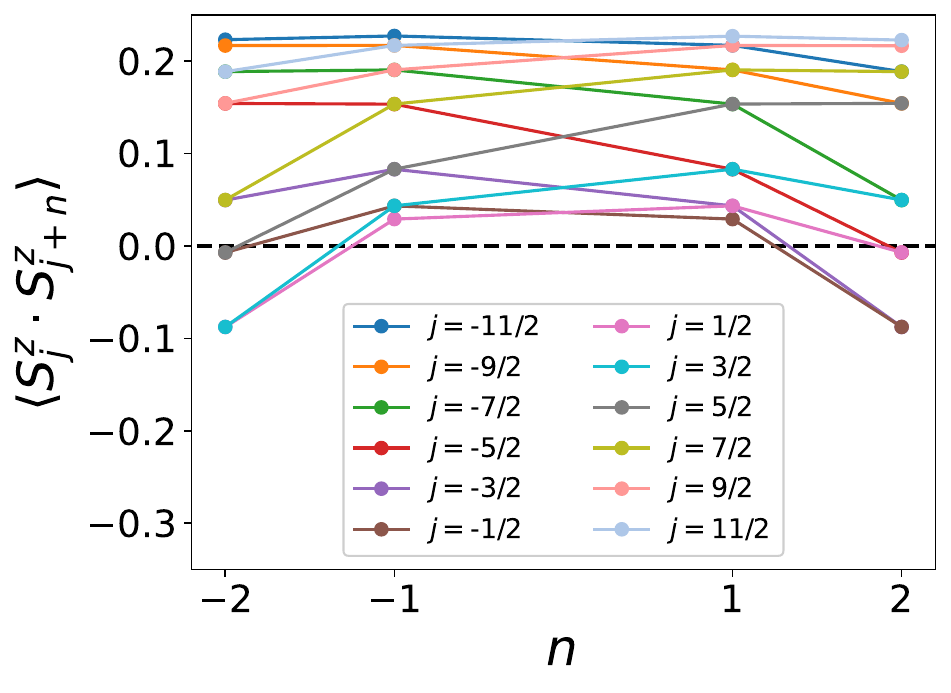}
        \caption{$\lambda_1=1.0$, $\lambda_2=1.2$}
        \label{AFMCorZZ}
    \end{subfigure}

    \caption{The local spin polarization $\braket{S^\mu_j} (\mu = x, y, z)$ and correlation functions 
$\braket{S^{\mu}_i S^{\mu}_j}$ near the impurity for $J_1/J_2=-0.5$ and $M=0.4$. 
The parameter are $\lambda_1=1.0$, and (a) (b) $\lambda_2=0.8$ and (c) (d) $\lambda_2=1.2$. }
    \label{Corall}
\end{figure*}

The characteristic behaviors of the magnetization around the impurity bond can be understood in more detail
based on spin correlations.
To this end,
let us generalize $\mathcal{H}_{\text{imp}}$ and introduce two parameters $\lambda_1,\lambda_2$, 
\begin{equation}
\begin{aligned}
    \mathcal{H}_{\text{imp}} &=  (\lambda_1-1) J_1 \bm{S}_{-\frac{1}{2}} \cdot \bm{S}_{\frac{1}{2}} \\
    &+ (\lambda_2-1) J_2 \left( \bm{S}_{-\frac{3}{2}} \cdot \bm{S}_{\frac{1}{2}} + \bm{S}_{-\frac{1}{2}} \cdot \bm{S}_{\frac{3}{2}} \right).
\end{aligned}
\label{eqimp2}
\end{equation}
By fixing one of  $\lambda_{1, 2}$ to $1$, we can discuss which interaction, $J_1$ or $J_2$, is dominant for the
magntetization near the impurity. Figure \ref{Corall} shows the local magnetization and correlation functions near the impurity for $\lambda_1=1$ and $\lambda_2 \neq 1$. 
The local magnetization is essentially similar to that with $\lambda_1=\lambda_2$ (Fig. \ref{magenall}).
We calculate the correlation function $\braket{S_j^{\mu}S_{j+n}^{\mu}}$ with $|n|\leq 2$ for each fixed site $j$.
For $\lambda_2<1$, the FM interaction $J_1$ is expected to be dominant over the AFM interaction
$J_2$ because of the reduction $J_2\to\lambda_2J_2$. 
Indeed, in Fig. \ref{Corall}, we can see enhanced FM correlations $\braket{S_j^zS_{j+n}^z}>0$ near the impurity bond
and suppressed AFM correlations in $x,y$-components. 
The FM correlation nearly reaches the maximum value, $\braket{\bm{S}_j\cdot\bm{S}_{j+n}}\simeq0.25$, 
for $j=\pm1/2,3/2$.
This corresponds to the saturation of the magnetization $\braket{S_j^z}\simeq0.5$. 
On the other hand, for $\lambda_2 >1$, the AFM interaction $J_2$ is dominant near the impurity. 
In fact, the correlation functions in Fig. \ref{Corall} show AFM behaviors in the very vicinity of the impurity,
which would correspond to spin-singlet pairs in the zero-field limit, $h_z\to0$. 
If spin-singlet pairs are formed and spins are fully frozen, they may be regarded as open boundaries for 
other spins and hence impurity effects for $\lambda>1$ would be similar to those for $\lambda<1$.
Indeed, the spin correlations become FM at sites slightly away from the impurity bond and becomes largest
at the sites $j \simeq \pm 11/2$. 
This corresponds to the local magnetization profile which is suppressed around $j \simeq \pm1/2$ and 
enhanced around $j \simeq \pm11/2$ (Fig. \ref{magenall}).

When $\lambda_2$ is fixed as $\lambda_2=1$, the AFM interaction $J_2$ is dominant for $\lambda_1<1$
and so is the FM interaction $J_1$ for $\lambda_1>1$.
We have confirmed that the same correspondence between the local magnetization and spin correlations holds
in this case as well.

\subsection{Friedel oscillations}
\label{sec:Friedel}

Here, we discuss the Friedel oscillations of the local magnetization $\braket{S_j^z}$ in detail.
According to the previous studies of a TLL with a local potential $V$~\cite{eggert1995, egger1995friedel, egger1996friedel}, 
the excess magntization $\delta S_j^z=\braket{S_j^z}-M$ measured from the bulk value $M$ 
behaves as 
\begin{align}
	\delta S_j^z \simeq \frac{A\cos (qj + \delta)}{j^{\eta}}\quad (j\gg 1), 
\label{eq:Friedel}
\end{align}
where $q=2\pi\rho=\pi (1/2-M)$ with the magnon-pair boson density $\rho$.
When the potential is weak, the exponent within a perturbation theory is given by 
\begin{align}
\eta = \left\{
\begin{array}{cc}
2K-1 & (j\ll j_0), \\
K & (j\gg j_0),
\end{array}
\right.
\end{align}
where $K$ is the TLL parameter. The crossover length scale $j_0$ depends on the potential strength $V$
and is larger (smaller) for a weaker (stronger) potential.
We find that it is difficult to identify $j_0$ in our numerical calculations of $\delta S_j^z$ and cannot clearly observe
the above two
distinct scaling behaviors.
Instead,
it turns out that 
$\delta S_j^z$ behaves as if it has an exponent $\eta$ in between the above two values,
$2K-1\leq \eta \leq K$, in the spin nematic phase.
(Note that $1/2<K<1$ in the spin nematic phase in the present notation~\cite{hikihara2008vector}.)
This makes fitting of the numerical results with Eq. \eqref{eq:Friedel} subtle to some extent, 
but we find that it works well and the fitting parameters $A, q, \delta$, and $\eta$ are numerically robust even for
a very weak potential ($\lambda\simeq 1$).
When the potential is not too weak, the Friedel oscillation obeys the standard algebraic behavior with 
$\eta=K$ for a wide region, and there is no difficulty in the fitting.
Thus, we suppose that the fitting parameters contain physical information of the system and
analyze their behaviors for a wide range of $\lambda$.

\begin{figure}[htb]
    
    \begin{subfigure}{0.23\textwidth}
        \includegraphics[width=\linewidth]{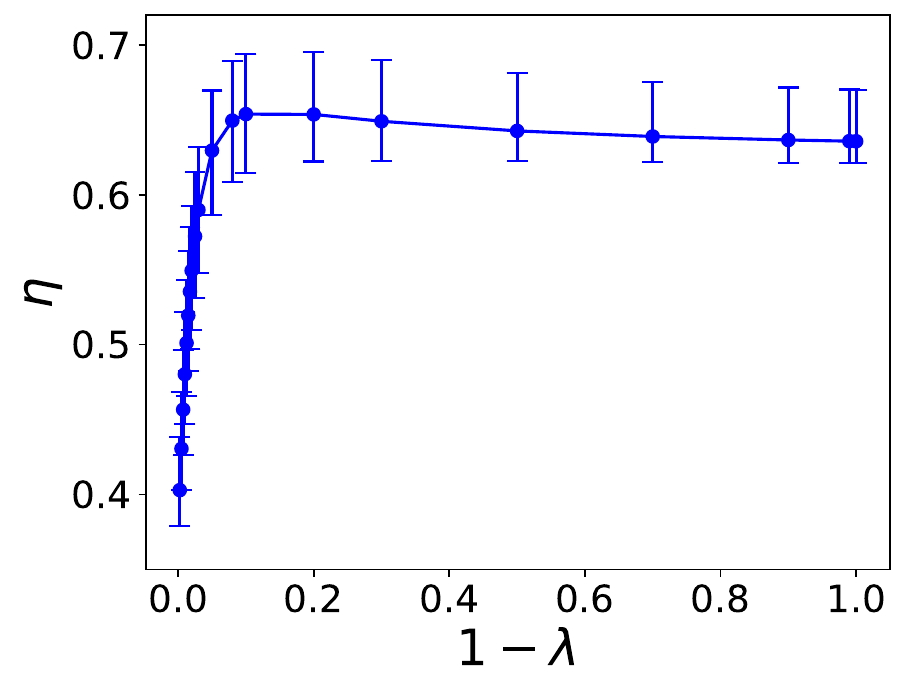}
        \caption{}
        \label{fig:Friedel1}
    \end{subfigure}
    \begin{subfigure}{0.23\textwidth}
        \includegraphics[width=\linewidth]{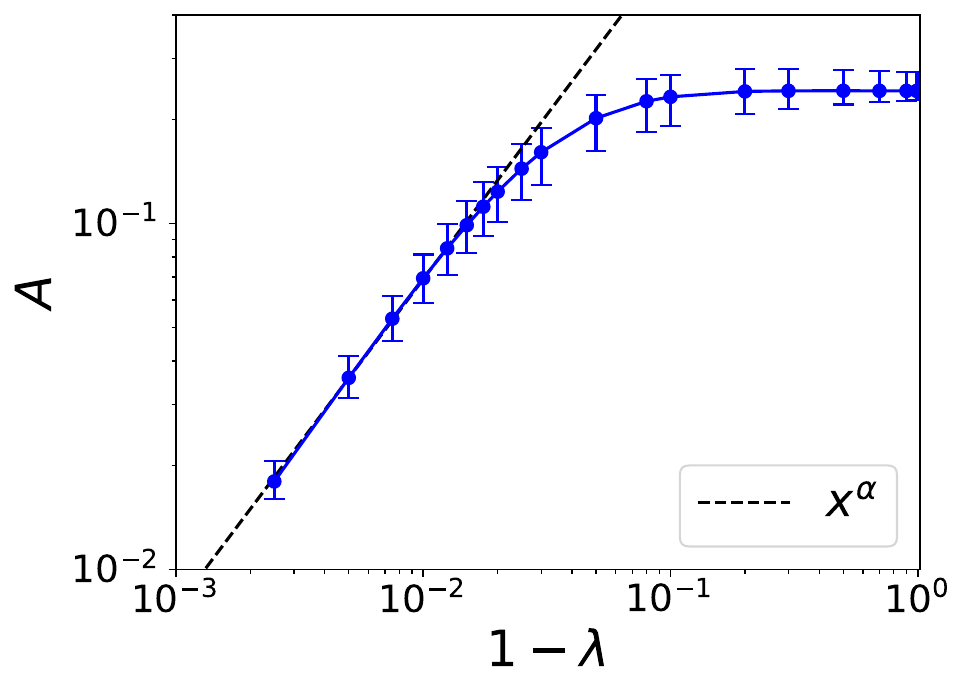}
        \caption{}
        \label{fig:Friedel2}
    \end{subfigure}

    \caption{(a) The exponent $\eta$ and (b) the amplitude $A$ of the Friedel oscillation of the 
excess magnetization $\delta S_j^z=\braket{S_j^z}_{\lambda}-\braket{S_j^z}_{\lambda=1}$ in the spin nematic phase.
The dashed line for the amplitude is $A\sim |1-\lambda|^{\alpha}$ with $\alpha= 0.95$.}
    \label{fig:Friedel}
\end{figure}

Figure \ref{fig:Friedel} shows the amplitude $A$ and the exponent $\eta$ obtained by fitting the numerical results 
of $\delta S_j^z$ for $\lambda<1$
with Eq.~\eqref{eq:Friedel}. 
Here, we have subtracted the local magnetization $\braket{S_j^z}|_{\lambda=1}$ as a background bulk magnetization.
The error bars correspond to different fitting regions.
The exponent is $\eta\simeq 0.4$ for very small values of $\delta\lambda=|1-\lambda|$, 
and it rapidly increases and converges
to a constant value $\eta\simeq 0.63=K_{\eta}$ 
which is (within the error bars) consistent with the bulk TLL parameter 
$K_{\rm bulk}\simeq 0.61$ estimated from
correlation functions for the uniform system with $\lambda=1$ (not shown).
The amplitude $A$ 
exhibits a scaling behavior $A\sim \delta\lambda^{\alpha}$ with an exponent $\alpha$ 
for $\delta\lambda\lesssim 0.1$, and saturates to a constant for $\delta\lambda\gtrsim 0.1$. 
This corresponds to the $\lambda$ dependence of the magnitude of the local magnetization around the impurity
$\braket{S_{\rm imp}^z}$ in Fig. \ref{magenall},
and the saturation of $A$ for large $\delta\lambda$ is simply because $|\braket{S_j^z}|\leq 1/2$.
The exponent $\alpha$ obtained from the fitting with the smallest four or five values of 
$\delta\lambda$ is $\alpha\simeq 0.95$,
and
it is related with the TLL parameter and is given by $\alpha=K/(2-2K)$ as will be discussed in the next section.
This leads to a TLL parameter defined from the amplitude, $K_{A}\simeq 0.65$, 
and it is roughly consistent with the TLL parameters $K_{\eta}, K_{\rm bulk}$ evaluated in the different ways.
We note that similar behaviors of the Friedel oscillations are expected for $\lambda>1$,
but detailed analyses are not easy due to the sharp dip near $j\sim0$.

In addition to the critical properties, the oscillation period $q^{-1}$ is also an important quantity.
We numerically evaluate the period to be $q\simeq 0.31$ from the fitting and it is consistent with the one estimated from the magnetization, $q=2\pi\rho=2\pi(1/2-M)/2=0.1\pi$. 
This value of $q$ is almost independent of the parameter $\lambda$ as expected.
The overall factor $1/2$ in the magnon pair density $\rho$ is a characteristic property in the spin nematic state
as discussed in the previous studies~\cite{hikihara2008vector},
and such a factor is absent in the standard TLL in magnetic fields (see Appendix \ref{app:XXZ}).
If the oscillation period $q^{-1}$ could be measured in an experiment, it would be strong evidence for that
elementary magnetic exciations in the spin nematic state are indeed magnon bound pairs but not single magnons.

\subsection{Bosonization argument}
\label{sec:Kane-Fisher}
We discuss 
the above numerical results of the local magnetization and entanglement entropy from a field theoretical point of view.
Here we consider the 
phenomenological description of the spin nematic state with nearly saturated uniform magnetization, $M\lesssim0.5$,
which provides a clear physical picture~\cite{kecke2007multimagnon, hikihara2008vector}.
The low energy physics in the spin nematic state are described by the two magnon bound pairs.
Theoretically, one can introduce hard-core bosons for magnon pairs 
\begin{equation}
\begin{aligned}
	b^{\dagger}_j\sim (-1)^jS_{j}^-S_{j+1}^-,\quad n_j=b^{\dagger}_jb_j\sim \frac{1}{2}\left(\frac{1}{2}-S_j^z\right),
\end{aligned}
\end{equation}
where $S_j^-=(S_j^x-iS_j^y)$ creates a single magnon at site $j$.
Then, the bosonization of an effective Hamiltonian for the magnon pair bosons in the bulk gives
\begin{equation}
\begin{aligned}
	{\mathcal H}_{\rm bulk}=\frac{v}{2}\int dx \left( K\left(\partial_x\theta\right)^2
	+\frac{1}{K}\left(\partial_x\phi\right)^2 \right),
\end{aligned}
\end{equation}
where $v$ is the velocity and 
$(\phi,\theta)$ are the conjugate field operators satisfying $[\phi(x),\partial_{x'}\theta(x')]=i\delta(x-x')$.
The TLL parameter is $1/2<K<1$ for the spin nematic phase and the bosonization is valid also in the SDW phase
with a relatively small magnetization for which the TLL parameter is $0<K<1/2$~\cite{hikihara2008vector}.

In presence of an impurity, there arise local perturbation terms corresponding to 
the spin model Hamiltonian ${\mathcal H}_{\rm imp}$.
For the present spin nematic system, it is a priori not clear how the lattice impurity term ${\mathcal H}_{\rm imp}$
is described in the effective boson picture.
Indeed, we cannot simply rewrite the spin model perturbation $\bm{S}_{i}\cdot\bm{S}_{j}$ in terms of 
the phenomenological degrees of freedom $b^{(\dagger)}_j$.
However, the local magnetization profile and spin correlations in the previous sections suggest that
${\mathcal H}_{\rm imp}$ leads to a repulsive potential for $\lambda<1$ near the impurity bond, 
since an increase (decrease) of local magnetization corresponds to a decrease (increase) of magnon pairs.
This implies that the impurity term with $\lambda<1$ is effectively described by
\begin{equation}
    \mathcal{H}_{\text{imp}} = V\sum_{j\sim0}n_j = V\sum_{j\sim0}\frac{1}{2}\left( \frac{1}{2} - S_{j}^z \right),
\label{Himp}
\end{equation}
where the summation is restricted to a few sites near the impurity bond.
The case with $\lambda>1$ is somewhat subtle because of the sharp dip of the magnetization around $j=\pm 1/2$.
The dip in $\braket{S_j^z}$ may imply that there is an attractive potential for magnon pairs,
but the numerical result $\braket{S_j^z}\simeq 0$ at the dip may be beyond the magnon pair picture which
is valid for a low magnon density.
Nevertheless, the saturation of the magnetization and the effective decoupling takes place at the sites $j_{\rm dec}$ 
slightly away from the impurity bond.
Therefore, we suppose that the effective boson description is valid for $|j|\gtrsim j_{\rm dec}$ and 
the impurity potential is repulsive also for $\lambda>1$.
Although it is not trivial how the phenomenological potential $V$ is related to the spin model parameter $\lambda$,
we expect that the relation
$V^2\propto |1-\lambda|= \delta\lambda$ holds. 
This is understood as an analogy of the basic fact that, in a general system, a Zeeman term generates
a quadratic spin-spin interaction in the second order perturbation.
There may be higher order perturbations in $b_j^{(\dagger)}$, 
but the lowest order term 
is usually most relevant and other terms would be less important.
For example, a local densinty-density interaction $\sim \delta\lambda n_jn_{j+1}$ around 
the impurity is expected to arises, but this term is irrelevant in the spin nematic state 
with the TLL parameter $1/2<K<1$ and cannot lead to effective decoupling of the system.
The relation $V\sim \sqrt{\delta\lambda}$ cannot be obtained by a standard perturbation expansion with 
respect to $\delta\lambda$ and is regarded as an ansatz. It will be supported by the consistency
between the field theory and the numerical calculations as discussed below.

\begin{figure}[htb]
	\includegraphics[width=0.33\textwidth]{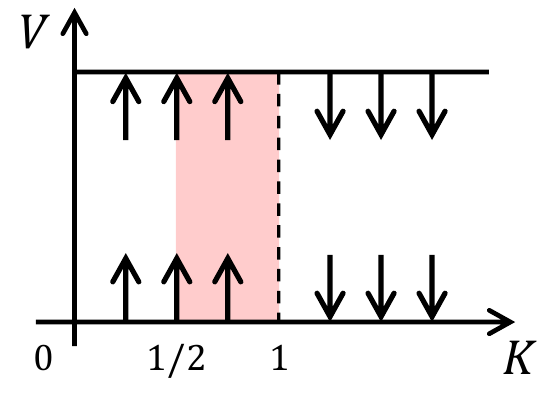}
	\caption{Renormalization group flow in the Kane-Fisher problem. In the spin nematic phase, the TLL parameter is $1/2<K<1$ (red region). }
	\label{RG}
\end{figure}

The total Hamiltonian ${\mathcal H}={\mathcal H}_{\rm bulk}+{\mathcal H}_{\rm imp}$ in the bosonized form is
\begin{align}
    \mathcal{H} &= \frac{v}{2}\int dx \left( K\left(\partial_x\theta\right)^2
	+\frac{1}{K}\left(\partial_x\phi\right)^2 \right) \nonumber\\ 
	&\quad +V\cos(\sqrt{4\pi}\phi(0)), 
\end{align}
where only the most important terms have been retained. 
It is known that the potential $V$ is effectively enhanced in the low energy regime of the renormalization group flow
for $K<1$ as shown in Fig. \ref{RG},
and correspondingly, the system is decoupled by the local perturbation~\cite{kane1992transport, kane1992resonant, kane1992transmission, furusaki1993resonant, furusaki1993single}.
Indeed, we have seen in Sec. \ref{sec:Magnetization} 
that the entanglement entropy around the impurity bond approaches zero
and the ground state wavefunction is essentially a product state of the left and right halves of the system. 
In this way, the numerical and field-theoretical approaches provide 
complementary understandings of the impurity effects.

One can also discuss how the local magnetization around the impurity $\braket{S_j^z}$ behaves 
based on the renormalization of the perturbation term ${\mathcal H}_{\rm imp}$.
The potential term ${\mathcal H}_{\rm imp}$ 
is renormalized as $\tilde{V}= V b^{1-K}$, 
or equivalently, the energy scale for $\tilde{V}=O(1)$ is given by 
$b^{-1}\sim V^{1/(1-K)}$~\cite{kane1992transport, kane1992resonant, kane1992transmission, furusaki1993resonant, furusaki1993single}.
Therefore, the local perturbation $\cos(\sqrt{4\pi}\phi(0))$ with the scaling dimension $K$ behaves as
\begin{align}
	\braket{\cos(\sqrt{4\pi}\phi(0))}\sim b^{-K}\sim V^{\frac{K}{1-K}} \sim \delta\lambda^{\frac{K}{2-2K}}
\end{align}
where the relation $V\sim \sqrt{\delta\lambda}$ has been used.
This implies that the local magnetization or equivalently the amplitude of the Friedel oscillation
$A$ in Eq.~\eqref{eq:Friedel} has the same scaling property, 
\begin{align}
	\braket{S_{j\sim0}^z}\sim A\sim \delta\lambda^{\frac{K}{2-2K}}.
\end{align}
As mentioned in Sec.~\ref{sec:Friedel}, this explains the qualitative behaviors of the numerically
obtained amplitude of the Friedel oscillation induced by the impurity.
The consistency between the field theory and the numerical calculations in turn
supports our ansatz that the phenomenological parameter and the spin model parameter satisfy
the relation $V\sim\sqrt{\delta\lambda}$.

\subsection{Discussion on NMR spectra}
\label{sec:NMR}
As was mentioned in Sec.~\ref{introduction}, 
it is difficult to observe the spin nematic order directly by the standard experimental techniques,
since the order parameter of the spin nematic state is the spin quadrupole moment defined by the product of two spin operators and there is no ordered dipole moment. 
However, our calculation results suggest that impurities can lead to characteristic magnetization profiles
in the spin nematic state, which could be observed by a local magnetization measurement such as NMR experiments.
Here, we discuss how the impurity effects can be seen in NMR spectra in the spin nematic phase.

\begin{figure}[htb]
    
    \begin{subfigure}{0.23\textwidth}
        \includegraphics[width=\linewidth]{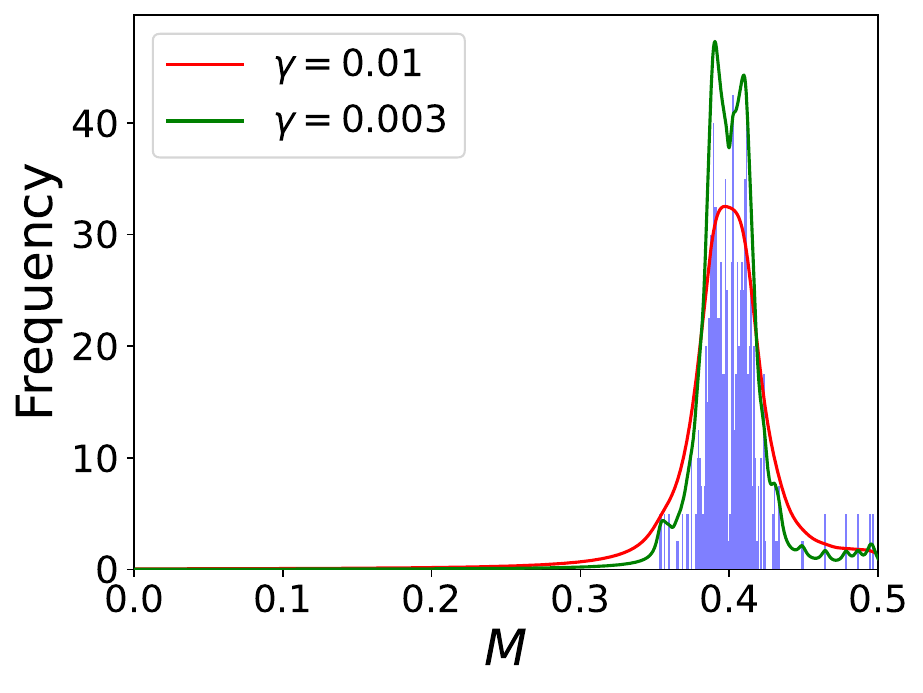}
        \caption{$\lambda=0.8$}
        \label{nmr1}
    \end{subfigure}
    \begin{subfigure}{0.23\textwidth}
        \includegraphics[width=\linewidth]{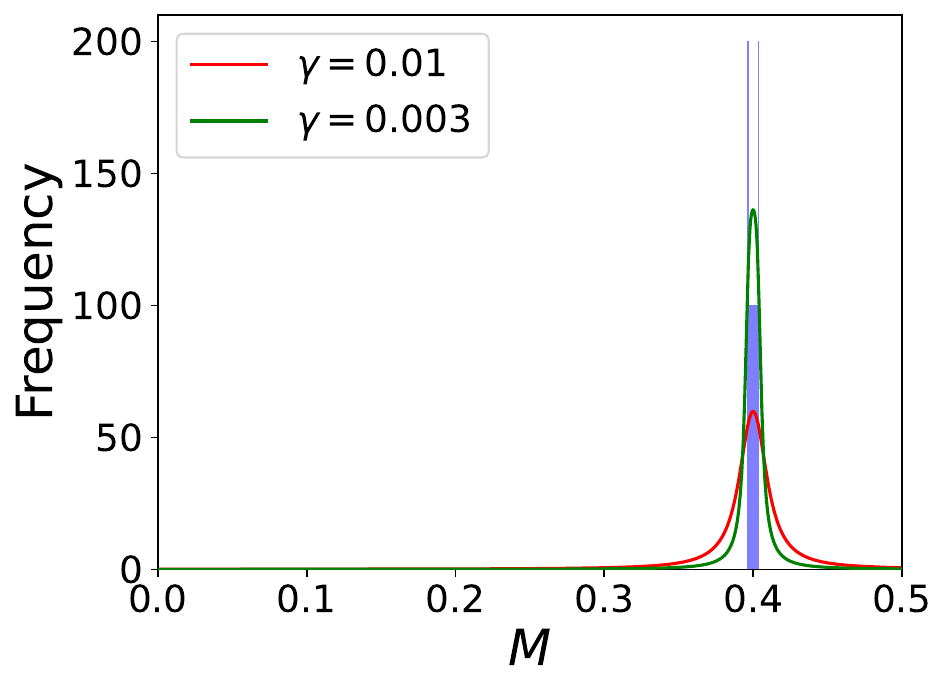}
        \caption{$\lambda=1.0$}
        \label{nmr2}
    \end{subfigure}
    \begin{subfigure}{0.23\textwidth}
        \includegraphics[width=\linewidth]{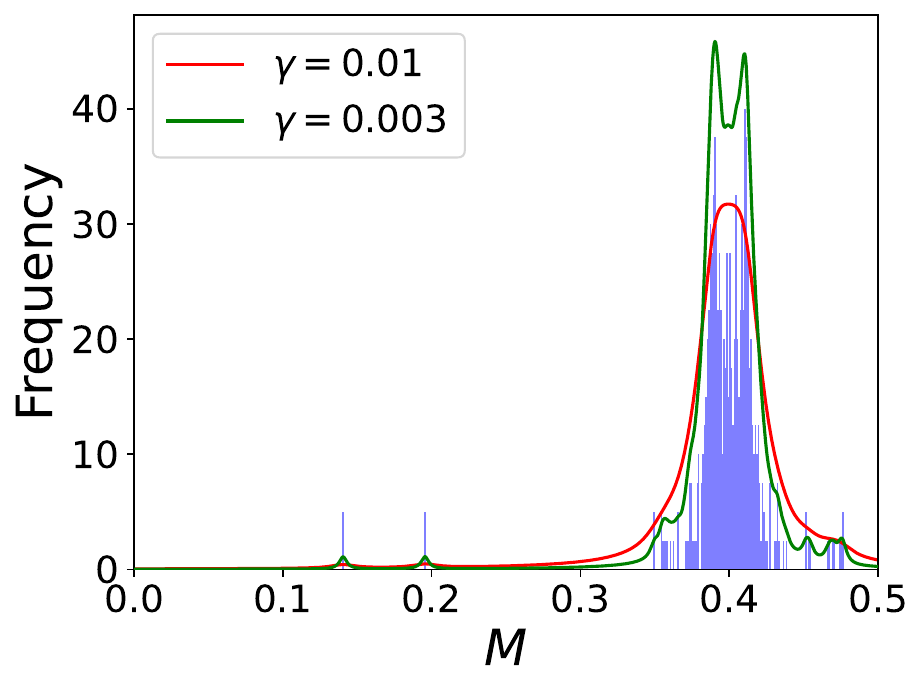}
        \caption{$\lambda=1.2$}
        \label{nmr3}
    \end{subfigure}
    \begin{subfigure}{0.23\textwidth}
        \includegraphics[width=\linewidth]{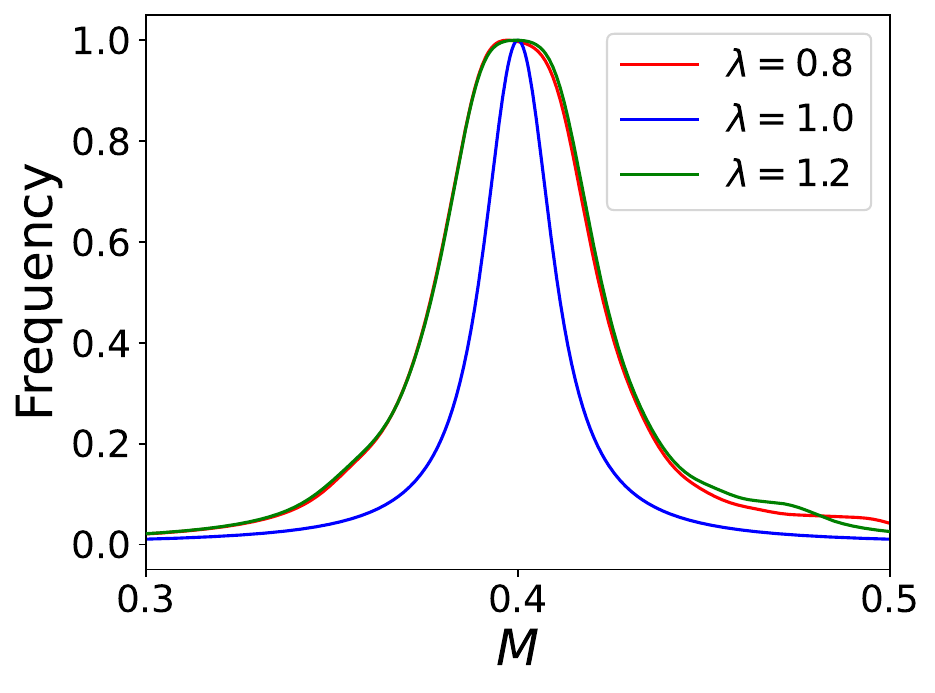}
        \caption{$\gamma=0.01$}
        \label{nmrshape}
    \end{subfigure}

    \caption{(a) (b) (c) Histograms of magnetization in the spin nematic phase with $J_1/J_2=-0.5$ and $M=0.4$
	for the $L=400$ window region. The curves are Lorentzian fitting with the phenomenological broadening 
parameter $\gamma$. (d) Normalized fitting curves with $\gamma=0.01$.}
    \label{nmrall}
\end{figure}

Figure \ref{nmrall} shows normalized histograms of the magnetization profiles in the window region $|j|<200$
corresponding to Fig.\ref{magenall} with Lorentzian fitting (see Appendix~\ref{app:Lorentzian} for details of the fitting). 
Roughly speaking, these are magnetization histrograms for systems with one impurity for every 400 sites,
although our model contains a single impurity bond in the infinite chain.
The histogram corresponds to an internal magnetic field observed as an NMR spectrum. 
For a uniform chain ($\lambda=1$) in the spin nematic phase,
the histogram has a single peak at the bulk magnetization $M=M_{\rm bulk}=0.4$.
The peak structure is smeared for $\lambda=0.8$, and importantly, there exists a reasonable weight around the saturated
magnetization, $M\simeq0.5$.
This is a direct consequence of the enhanced magnetization near the impurity bond (Fig.\ref{magenall}).
The relative magnitude of this weight will be increased in a system with a larger impurity concentration.
In case of $\lambda=1.2$, there are small weights around $M=0.1\sim0.2$ in addition to the weight around $M=0.5$,
which corresponds to the sharp dip in the magnetization profile near the impurity bond.
In both cases, the impurity contributions are well separated from the bulk peaks localized at $M=M_{\rm bulk}$.
Besides, 
the line shapes of the histograms (Fig.~\ref{nmrshape}) in the spin nematic 
are rather sensitive to the impurity compared to those in the standard XXZ model at the same magnetization $M$
(see Appendix \ref{app:XXZ}).
This is basically because the TLL parameter for the spin nematic state is smaller than that of the XXZ model
for a fixed $M$, and the Friedel oscillation has a longer tail in the former.

\begin{figure}[htb]

    \begin{subfigure}{0.23\textwidth}
        \includegraphics[width=\linewidth]{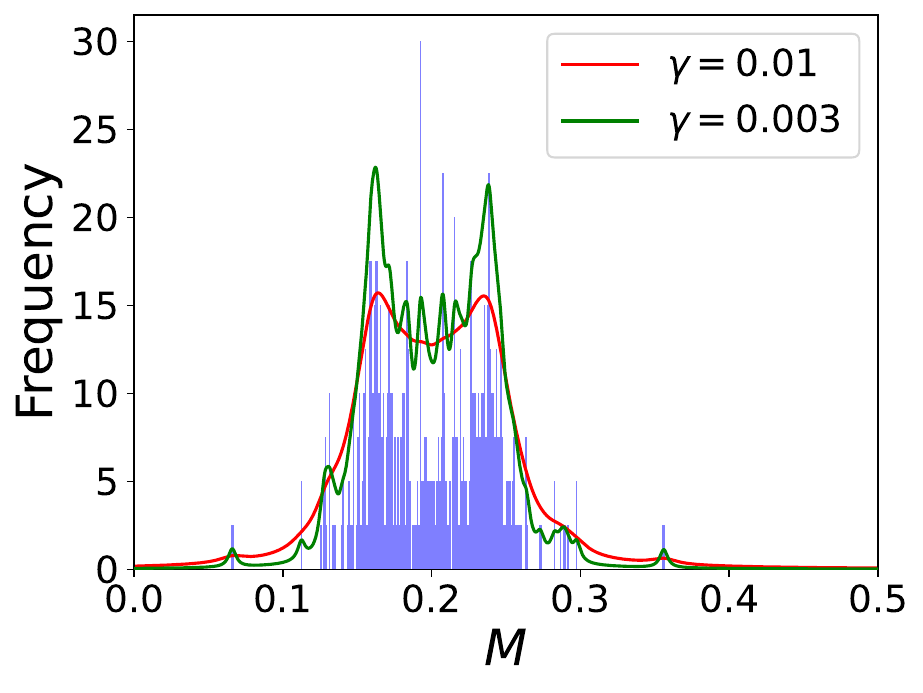}
        \caption{$\lambda=0.8$}
        \label{SDWnmr1}
    \end{subfigure}
    \begin{subfigure}{0.23\textwidth}
        \includegraphics[width=\linewidth]{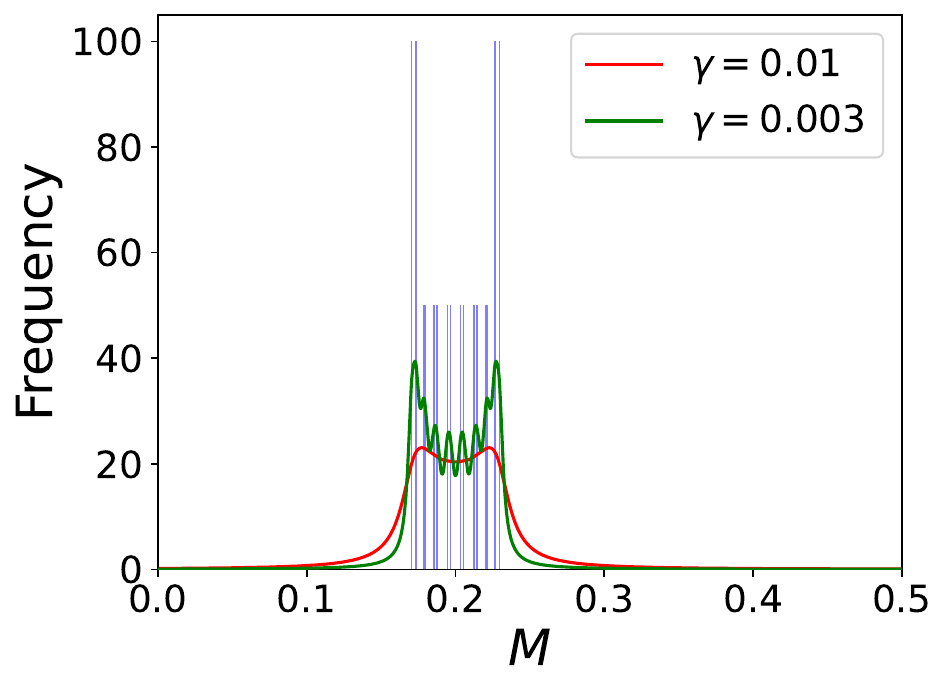}
        \caption{$\lambda=1.0$}
        \label{SDWnmr2}
    \end{subfigure}
    \begin{subfigure}{0.23\textwidth}
        \includegraphics[width=\linewidth]{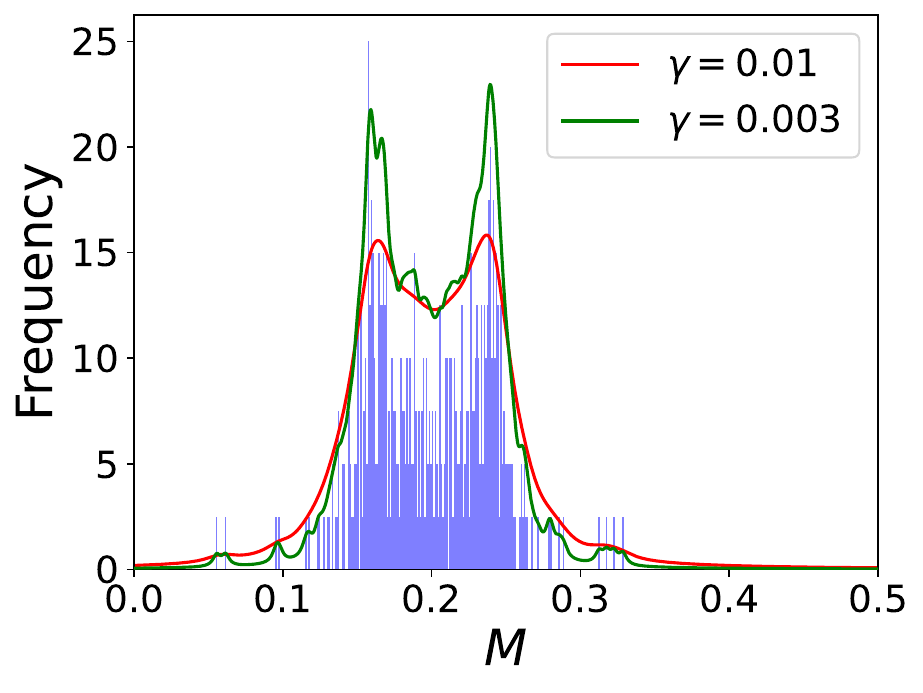}
        \caption{$\lambda=1.2$}
        \label{SDWnmr3}
    \end{subfigure}
    \begin{subfigure}{0.23\textwidth}
        \includegraphics[width=\linewidth]{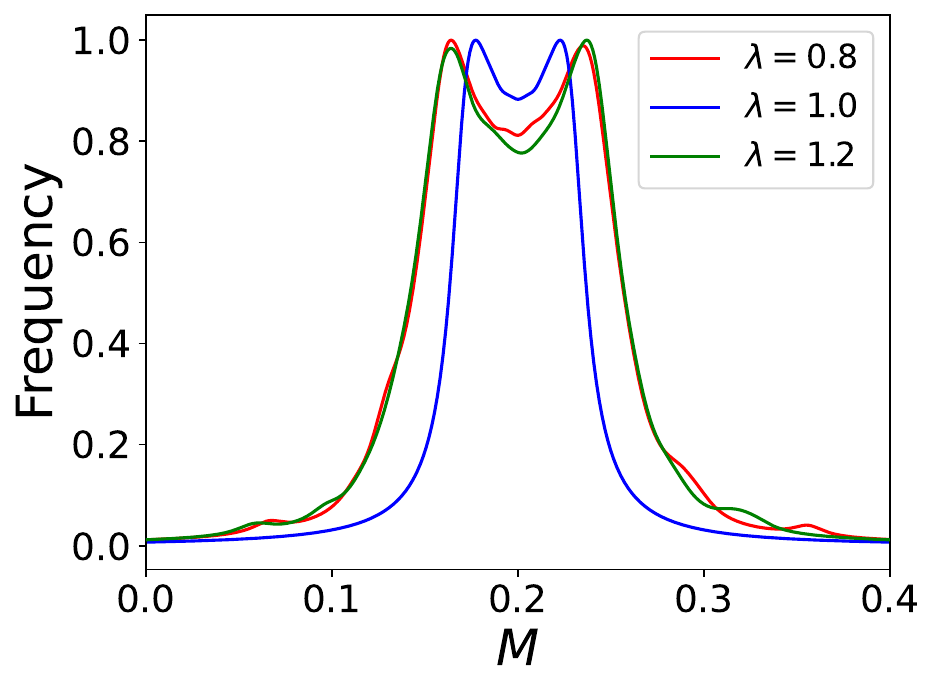}
        \caption{$\gamma=0.01$}
        \label{nmrshape}
    \end{subfigure}
    
    \caption{(a) (b) (c) Histograms of magnetization in the SDW phase with $J_1/J_2=-0.5$ and $M=0.2$ 
	for the $L=400$ window region. The curves are Lorentzian fitting with the phenomenological broadening 
parameter $\gamma$. (d) Normalized fitting curves with $\gamma=0.01$. }
    \label{SDWnmrall}
\end{figure}

Figure \ref{SDWnmrall} presents histograms of the magnetization corresponding to Fig.\ref{SDWmagenall} in the
SDW phase with the bulk moment $M_{\rm bulk}=0.2$. 
In this case, we can observe a double-horn shape of the histogram which is characteristic of  an SDW state. 
The double-horn structure is somewhat broadened for $\lambda\neq1$, but there is no qualitative difference 
between the results for $\lambda=0.8$ and $\lambda=1.2$.
Although there are some small contributions around $M=0.1$ and $M=0.3$, they are not clearly separated from
the main double-horn peaks.

We also consider magnetization histograms for the different impurity model which is closely related to
the previous NMR experiment \cite{buttgen2014search}.
Similar impurity contributions are found in this case as well (see Appendix \ref{app:defect} for details).
However, we should be careful when discussing real materials based on our model calculation.
For example, the NMR spectra at the V-sites were observed in the previous NMR studies in LiCuVO$_4$
~\cite{buttgen2014search,orlova2017},
while our histograms correspond to local magnetic fields at the Cu-sites.
The local magnetic fields at the V-sites would be smeared compared to those at the Cu-sites.


\section{Summary and discussion}
\label{sec:Summary}
We have discussed the impurity effects in the spin nematic state.
The iDMRG calculations were performed on  
the $J_1$-$J_2$ frustrated spin chain with $S=1/2$, where the impurities are modeled by 
the local change of the interaction strength characterized by the parameter $\lambda$. 
We found that for $\lambda< 1$, the magnetization near the impurity
nearly saturates, while for $\lambda>1.0$,
it has a sharp dip in the very vicinity of the impurity and tends to saturates at sites 
slightly away from the impurity. 
In both cases,
the entanglement entropy nearly vanishes around the impurity and the spin chain is effectively decoupled.
The local magnetization near the impurity can be further understood  
based on spin correlations and the Friedel oscillations exhibit characteristic critical behaviors. 
Our numerical results can be well understood based on
the phenomenological bosonization theory in presence of an impurity. 
The impurity induced local magnetization leads to additional contributions to the magnetization histogram.

In the present study, we focused on the single impurity problem and it can be a starting point for many-impurity
problems.
If there are multiple impurities which are spatially separated, 
the spin nematic order will be suppressed locally around each impurity. 
One may expect that impurities with a finite density will eventually suppress the global nematic order and
a disorder induced state could arise.
In such a disordered system, bulk quantities such as the magnetization curve and the steps
in $M(h_z)$ would be affected by the impurities.
It is also interesting to investigate how the finite density impurities will influence the binding energy of the
magnon pairs~\cite{Parvej2017}.
The present study could provide a basis for further understandings of impurities and disorder
in spin nematic states.

\begin{acknowledgments}
The numerical calculations were performed with the use of the TeNPy Library. 
We thank Masaki Oshikawa for valuable discussions especially on the bosonization analysis.
We are also grateful to  Shunsuke C. Furuya, Masashi Takigawa, and Hirokazu Tsunetsugu 
for constructive discussions and comments. 
This work is supported by JSPS KAKENHI Grant No. 22K03513.
\end{acknowledgments}


\appendix


\section{Effects of finite bond dimension and finite window size}
\label{app:finite_size}

As shown in the main text, the local magnetization and the entanglement entropy (i) have small oscillations
even at $\lambda=1$ and (ii) are not smoothly connected to those in the semi-infinite environments.
To understand these finite size effects, we perform numerical calculations for larger bond dimension $\chi$ 
and window sizes $L$ up to $\chi=1000$ and $L=1200$. 
It is found that the oscillation is related to the bond dimension and 
the discontinuity is to the window size, and they are suppressed for larger $\chi$ and $L$.

\begin{figure}[htb]
    \begin{subfigure}{0.23\textwidth}
        \includegraphics[width=\linewidth]{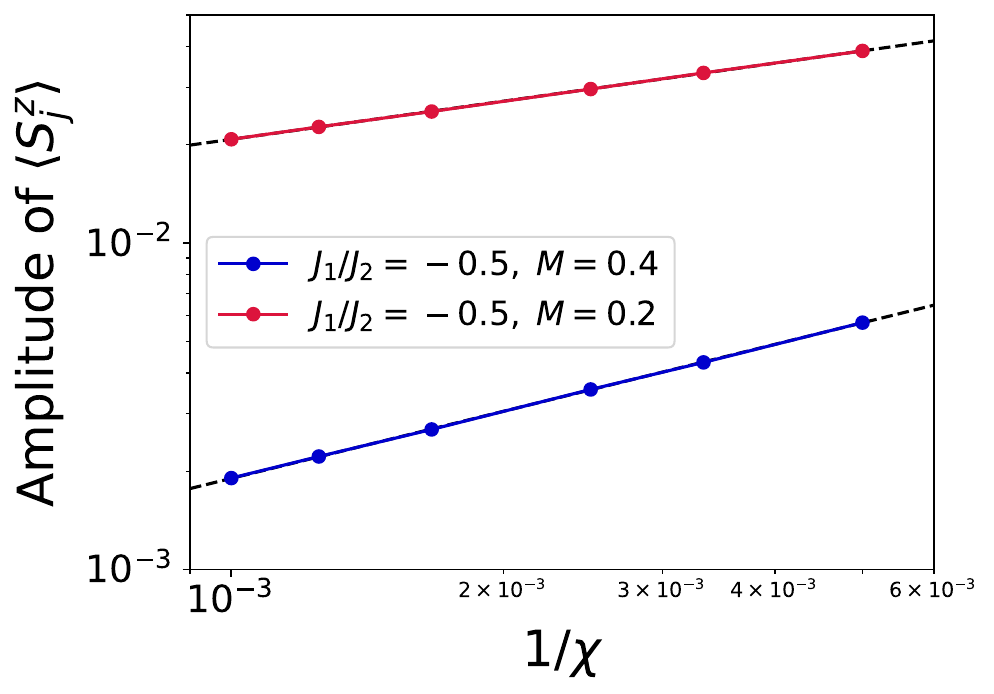}
        \caption{}
        \label{magLarge1}
    \end{subfigure}
    \begin{subfigure}{0.23\textwidth}
        \includegraphics[width=\linewidth]{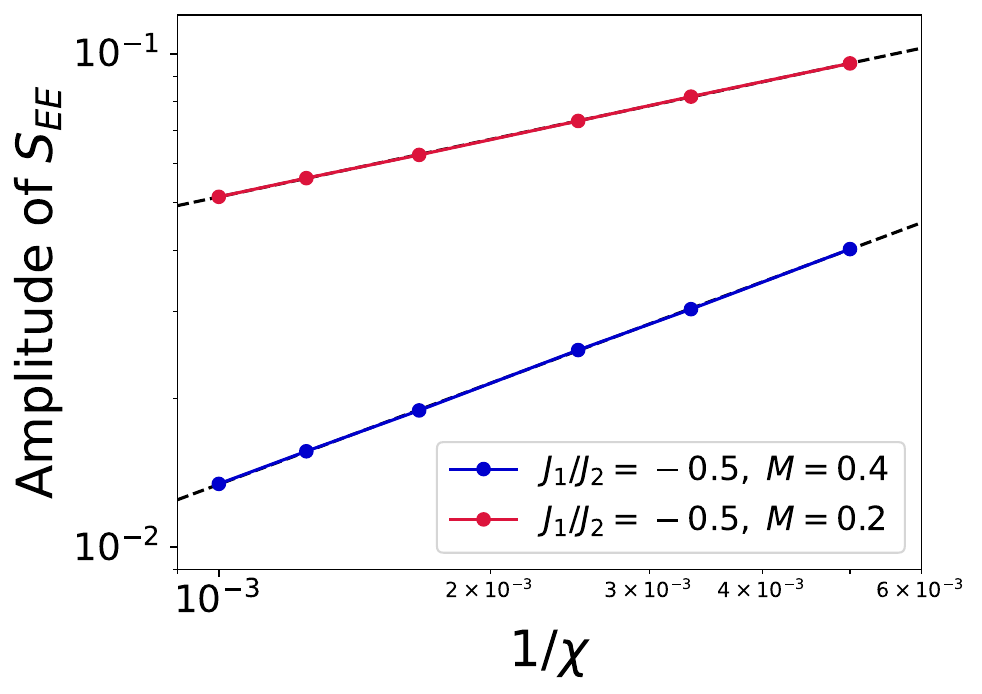}
        \caption{}
        \label{magsmall}
    \end{subfigure}
    
    \caption{Oscillation amplitudes of (a) the magnetization and (b) the entanglement entropy
for the uniform chain with $L=400$. 
The average magnetization $M=0.2$ corresponds to the SDW state and $M=0.4$ to the spin nematic state.}
    \label{Oscillation}
\end{figure}

We first discuss $\chi$-dependence of the oscillation amplitudes in the magnetization $(\Delta \braket{S^z})$ 
and the entanglement entropy $(\Delta S_{\rm EE})$
for a uniform chain with $L=400$. As seen in Fig.~\ref{Oscillation},
the oscillation amplitudes exhibit power law behaviros, $\Delta \braket{S^z}\sim \chi^{-\kappa_{\rm spin}}$ and 
$\Delta S_{\rm EE}\sim \chi^{-\kappa_{\rm EE}}$. 
They vanish for sufficiently large bond dimensions, $\chi\to\infty$, 
which means that the translation symmetry is not spotaneously broken
in the ground state of an infinite system.
However, 
the exponents are $\kappa_{\rm spin},\kappa_{\rm EE}<1$ and the oscillations remain even for large bond dimensions.
(More precisely, numerical fitting gives 
$\kappa_{\rm spin}=\kappa_{\rm EE}\simeq 0.68$ for the spin nematic state at $M=0.4$ and
$\kappa_{\rm spin}=\kappa_{\rm EE}\simeq 0.39$ for the SDW state at $M=0.2$.)
Neverthelss, overall behaviors are well described with a moderate bond dimension as discussed in the main text.

\begin{figure}[htb]

    \begin{subfigure}{0.23\textwidth}
        \includegraphics[width=\linewidth]{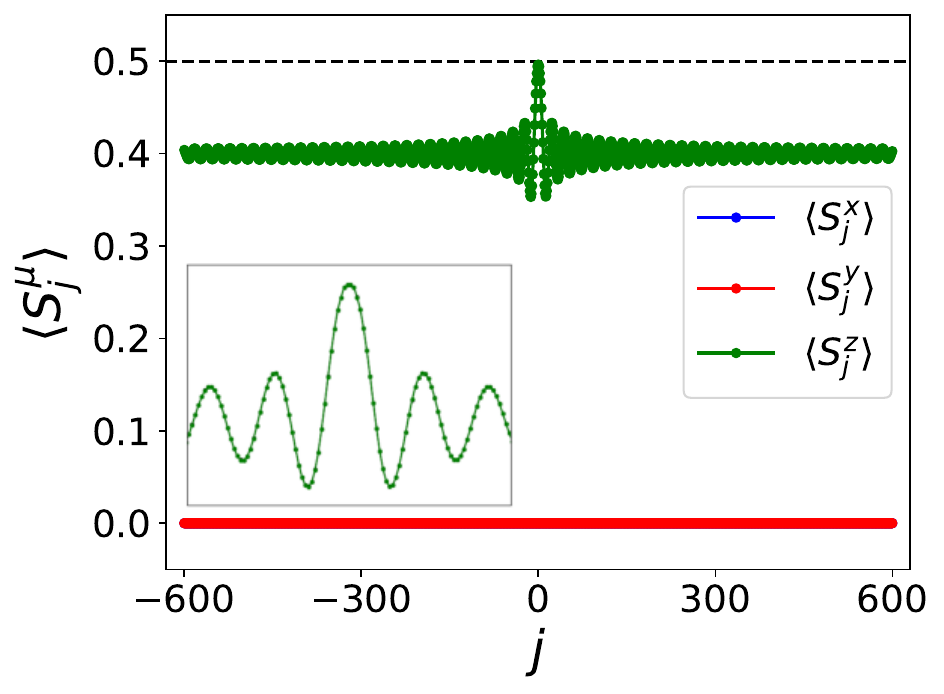}
        \caption{$\lambda=0.8$}
        \label{magLarge1}
    \end{subfigure}
    \begin{subfigure}{0.23\textwidth}
        \includegraphics[width=\linewidth]{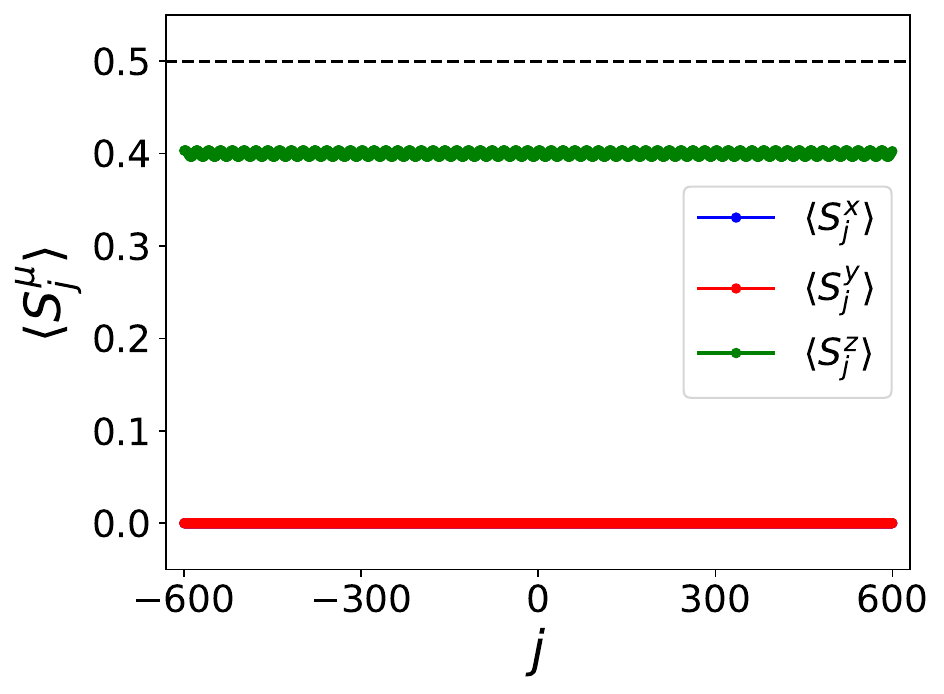}
        \caption{$\lambda=1.0$}
        \label{magsmall}
    \end{subfigure}
    \begin{subfigure}{0.23\textwidth}
        \includegraphics[width=\linewidth]{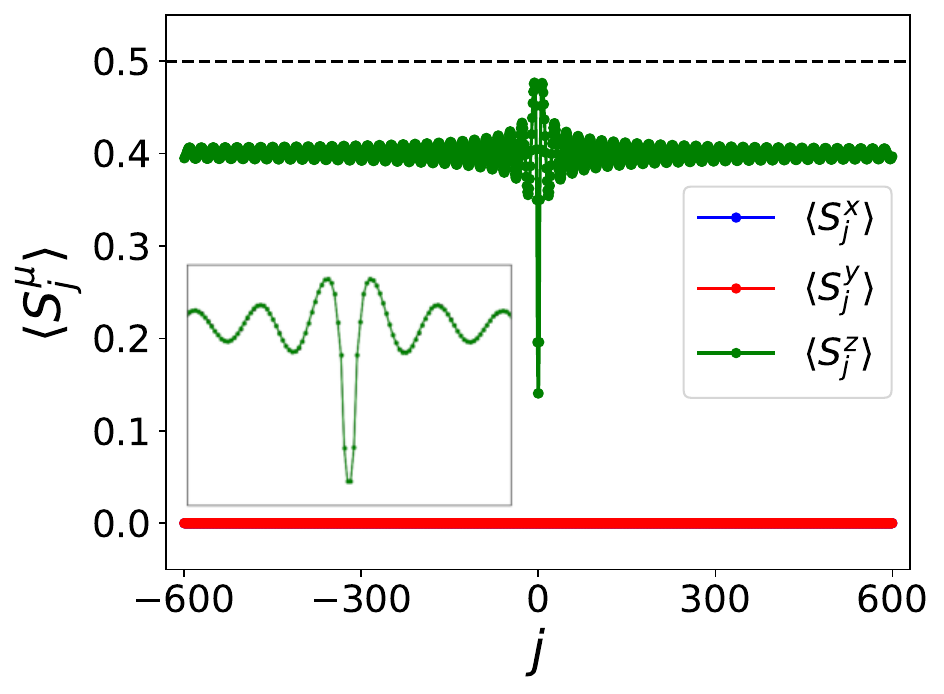}
        \caption{$\lambda=1.2$}
        \label{magLarge2}
    \end{subfigure}
    \begin{subfigure}{0.23\textwidth}
        \includegraphics[width=\linewidth]{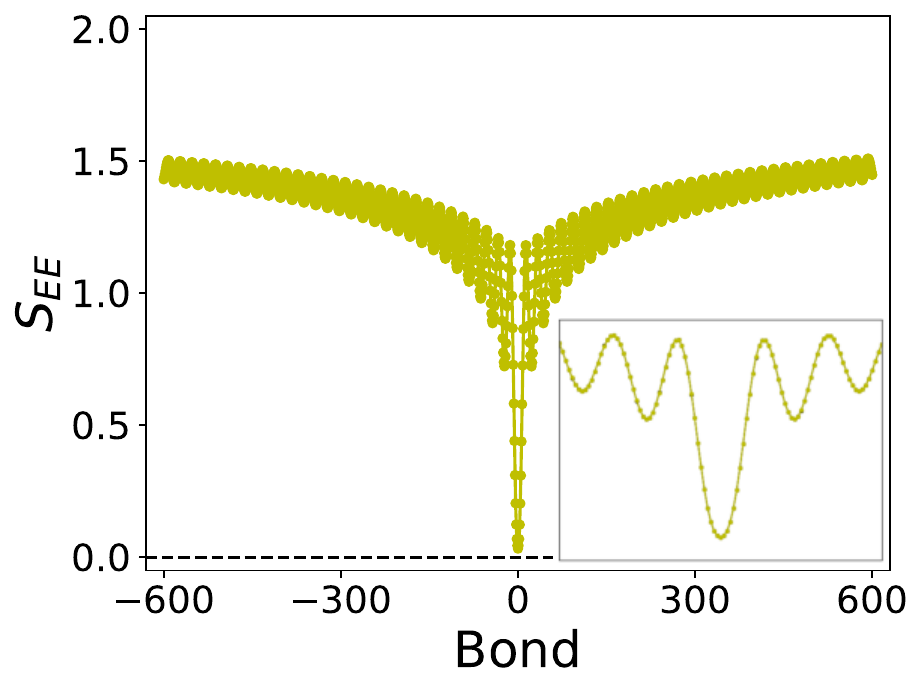}
        \caption{$\lambda=0.8$}
        \label{enLarge1}
    \end{subfigure}
    \begin{subfigure}{0.23\textwidth}
        \includegraphics[width=\linewidth]{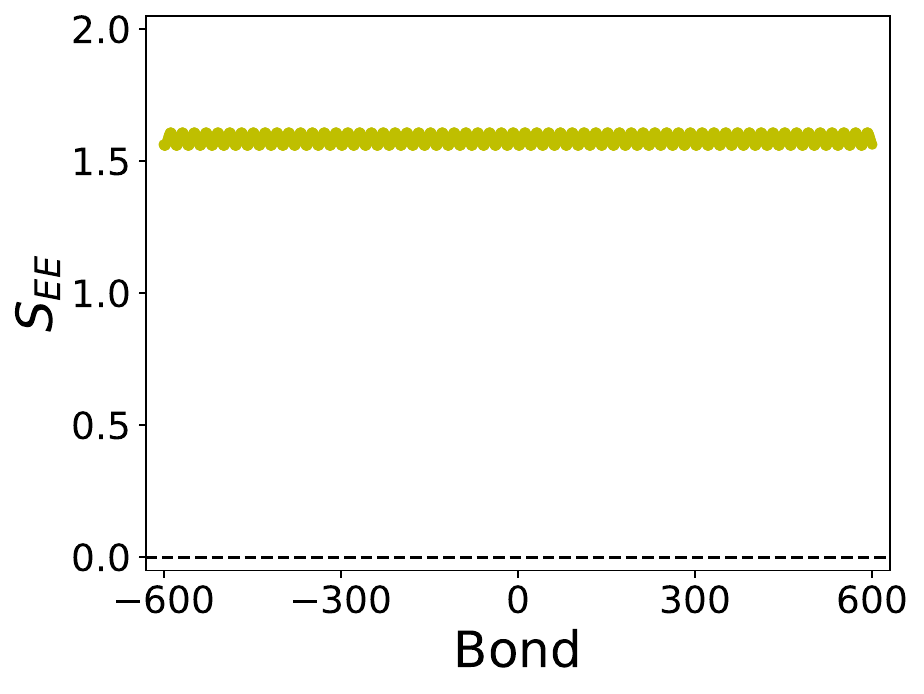}
        \caption{$\lambda=1.0$}
        \label{ensmall}
    \end{subfigure}
    \begin{subfigure}{0.23\textwidth}
        \includegraphics[width=\linewidth]{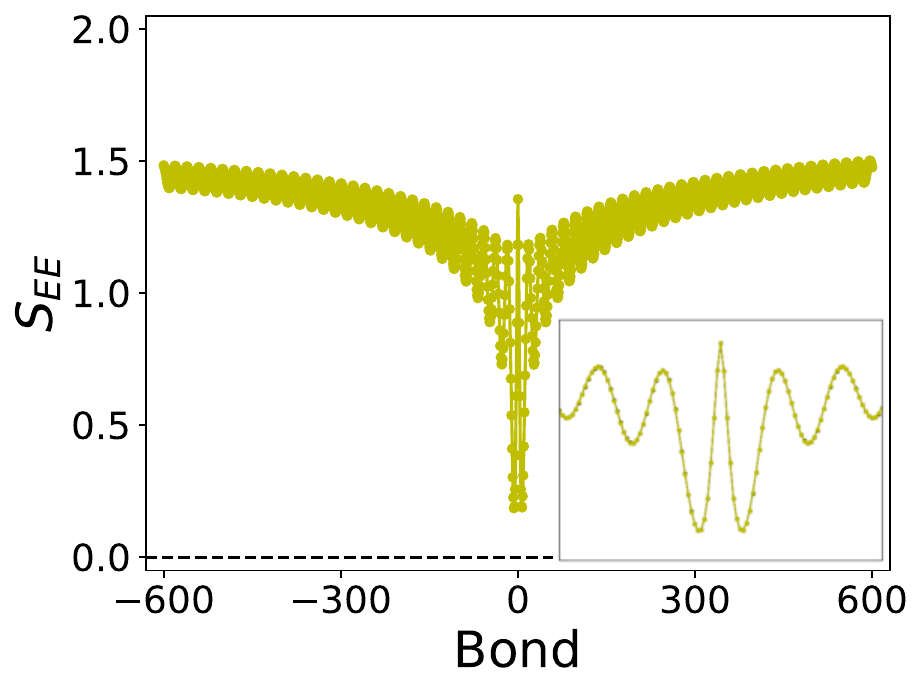}
        \caption{$\lambda=1.2$}
        \label{enLarge2}
    \end{subfigure}
    \caption{(a) (b) (c) The local spin polarization $\braket{S^\mu_j} (\mu = x, y, z)$ and (d) (e) (f) 
the entanglement entropy $S_{\text{EE}}$ for $L=1200$ in the spin nematic phase with $J_1/J_2=-0.5$ and $M=0.4$. 
The insets compare the results for $L=400$ and $L=1200$ with a focus on $100$ sites around the impurity,
where the two data conincide. }
    \label{Largeall}
\end{figure}

We next consider the discontinuity of the entanglement entropy at the boundaries of 
the window region and semi-infinite environments.
Figure \ref{Largeall} shows the numerical results for $L=1200$ with $\chi=400$.  
The oscillation amplitudes of $\braket{S_j^z}$ and $S_{EE}$ are the same as those for $L=400$ with $\chi=400$ in 
Fig. \ref{magenall} in the main text.
On the other hand,
from Figs.\ref{enLarge1}, \ref{ensmall}, and \ref{enLarge2}, we can see that the entanglement entropy is smoothly connected to the uniform system as the window size is increased. 
The insets show the comparison of the numerical results for $L=400$ and $L=1200$, focusing on 
the neighborhood of the impurity. 
One can see that the two curves for the different $L$ almost coincide and the finite window size effects
are negligible in the widonw region.

\section{Defect Model}
\label{app:defect}

\begin{figure}[htb]
	\includegraphics[width=0.37\textwidth]{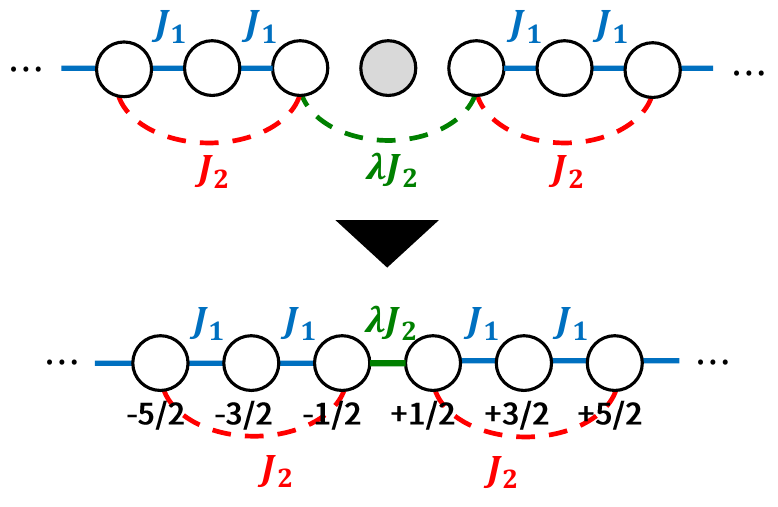}
	\caption{Schematic picture of the defect model. The parameter $\lambda$ characterizes the magnitude of the exchange interaction near the defect. 
The defect is a non-magnetic site and thus the top figure is reduced to the bottom figure.}
	\label{set2}
\end{figure}

In the main text, we have discussed the impurity system where the impurities locally change the magnitude of the exchange interaction. We can also consider other types of impurity systems, such as lattice defects and atomic substitutions. Here, we discuss the defect model introduced 
in the previous NMR study \cite{buttgen2014search}.
In this model, the spin at a defect site is frozen and does not interact with other spins. Therefore, we consider the impurity system without the defect site, as shown in Fig.\ref{set2}. The effective Hamiltonian is
${\mathcal H}=\mathcal{H}_{\text{bulk}}+\mathcal{H}_{\text{imp}},$
\begin{align}
	\mathcal{H}_{\text{bulk}} &=  J_1\sum_{j\neq -1/2} \bm{S}_j \cdot \bm{S}_{j+1} 
	+ J_2\sum_{j\neq -3/2,-1/2} \bm{S}_j \cdot \bm{S}_{j+2} \nonumber\\ 
	&\quad - h_z \sum_{j} S^{z}_j,\\
	\mathcal{H}_{\text{imp}} &=  \lambda J_2 \bm{S}_{-\frac{1}{2}} \cdot \bm{S}_{\frac{1}{2}}.
\end{align}

\begin{figure}[htb]
    \begin{subfigure}{0.23\textwidth}
        \includegraphics[width=\linewidth]{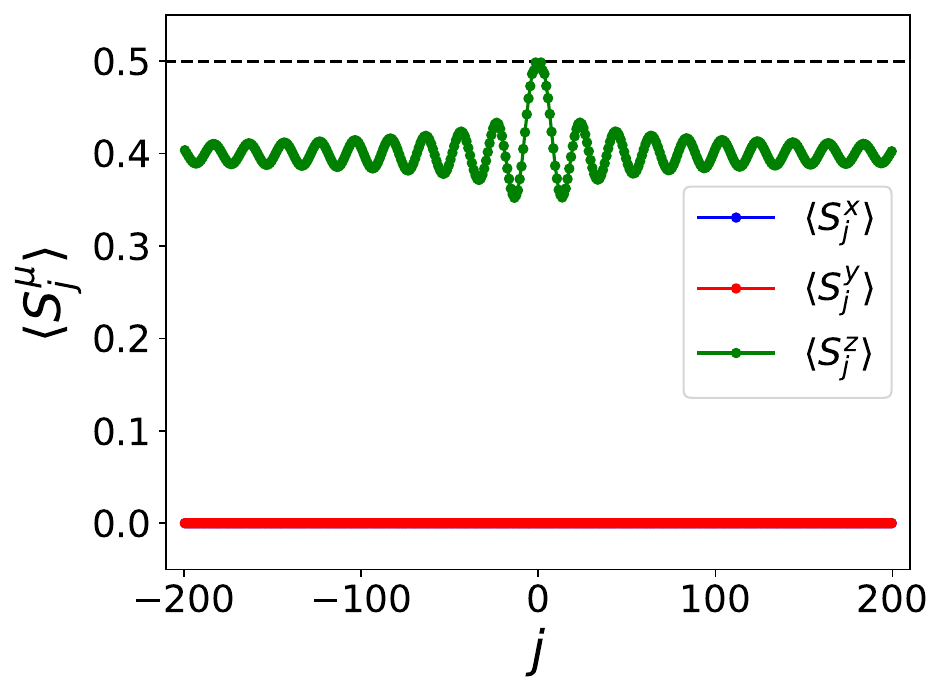}
        \caption{$\lambda=0.8$}
        \label{defectmag1}
    \end{subfigure}
    \begin{subfigure}{0.23\textwidth}
        \includegraphics[width=\linewidth]{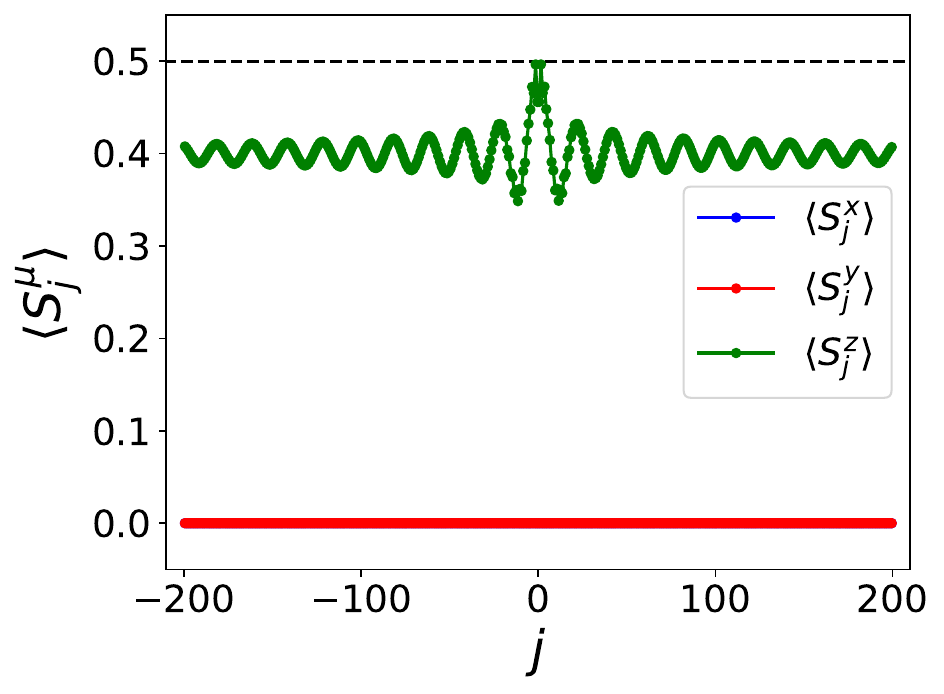}
        \caption{$\lambda=1.0$}
        \label{defectmag2}
    \end{subfigure}
    \begin{subfigure}{0.23\textwidth}
        \includegraphics[width=\linewidth]{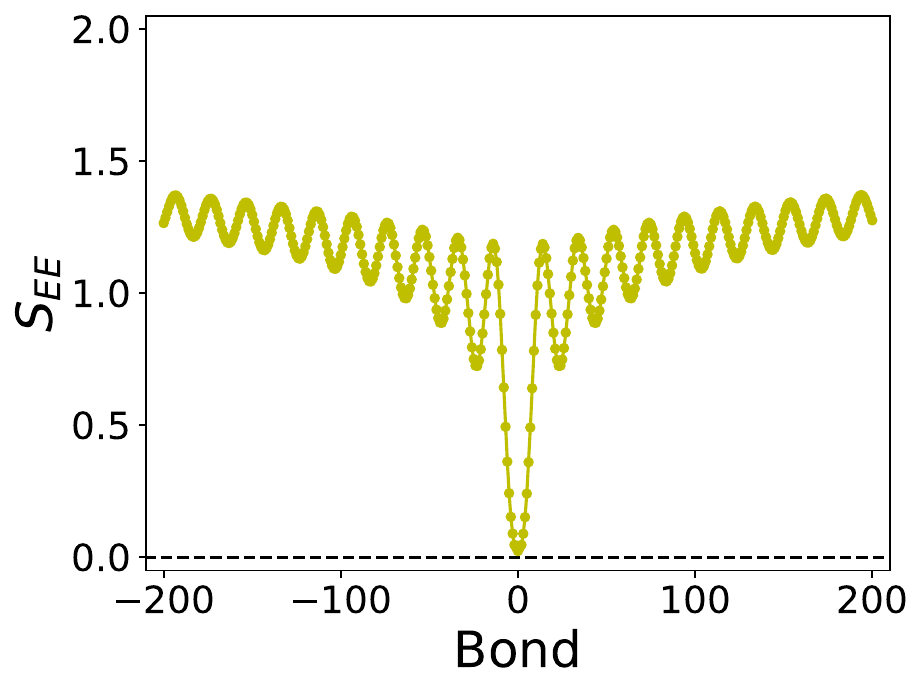}
        \caption{$\lambda=0.8$}
        \label{defecten1}
    \end{subfigure}
    \begin{subfigure}{0.23\textwidth}
        \includegraphics[width=\linewidth]{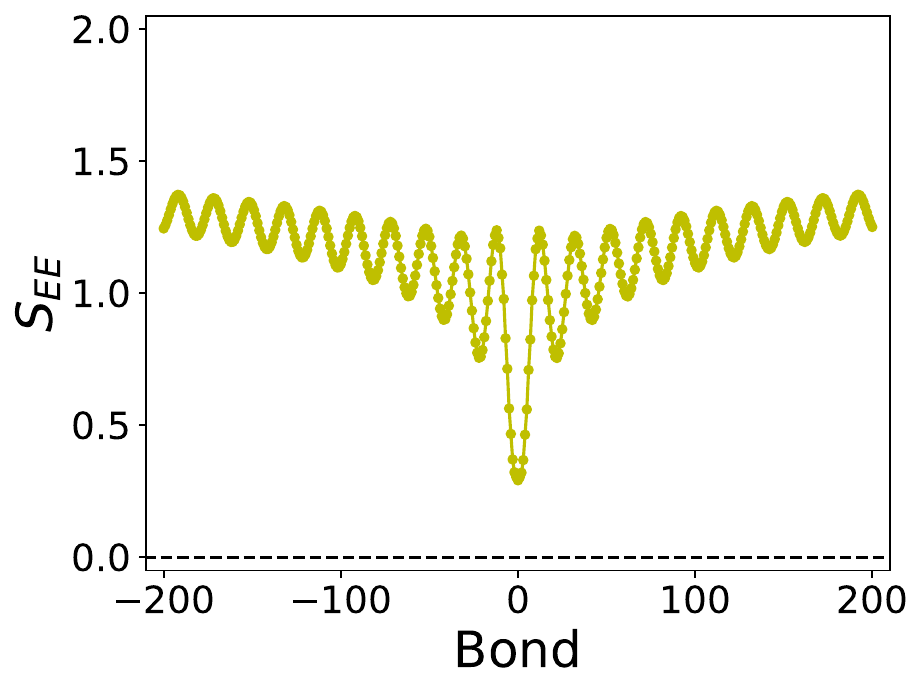}
        \caption{$\lambda=1.0$}
        \label{defecten2}
    \end{subfigure}
    
    \caption{(a) (b) (c) The local spin polarization $\braket{S^\mu_j} (\mu = x, y, z)$ and 
(d) (e) (f) the entanglement entropy $S_{\text{EE}}$ for $L = 400$ with $J_1/J_2=-0.5$ and $M=0.4$ in the defect model. }
    \label{defectmagenall}
\end{figure}

Figure \ref{defectmagenall} shows the local magnetization and the entanglement entropy in the spin nematic phase
at $J_1/J_2=-0.5$ and $M=0.4$. 
There are impurity contributions even for $\lambda=1$ since there is no translation symmetry in the present 
defect model.
For $\lambda=0.8$, $\braket{S_j^z}$ almost saturates around $j\sim0$ and correspondingly $S_{EE}$ is strongly suppressed.
Unfortunately, however, it is difficult to obtain good numerical convergence for $\lambda>1$.
Nevertheless, the poorly converged results show a sharp dip near the impurity bond 
and enhancement of the magnetization at sites slightly away from the impurity bond similarly to 
the results discussed in the main text.
Thus, similarly to the bond disorder model Eq. \eqref{eqimp1},
the effective decoupling of the spin chain around impurity takes place for both $\lambda<1$ and $\lambda>1$
also in the defect model.

\begin{figure}[htb]
    
    \begin{subfigure}{0.23\textwidth}
        \includegraphics[width=\linewidth]{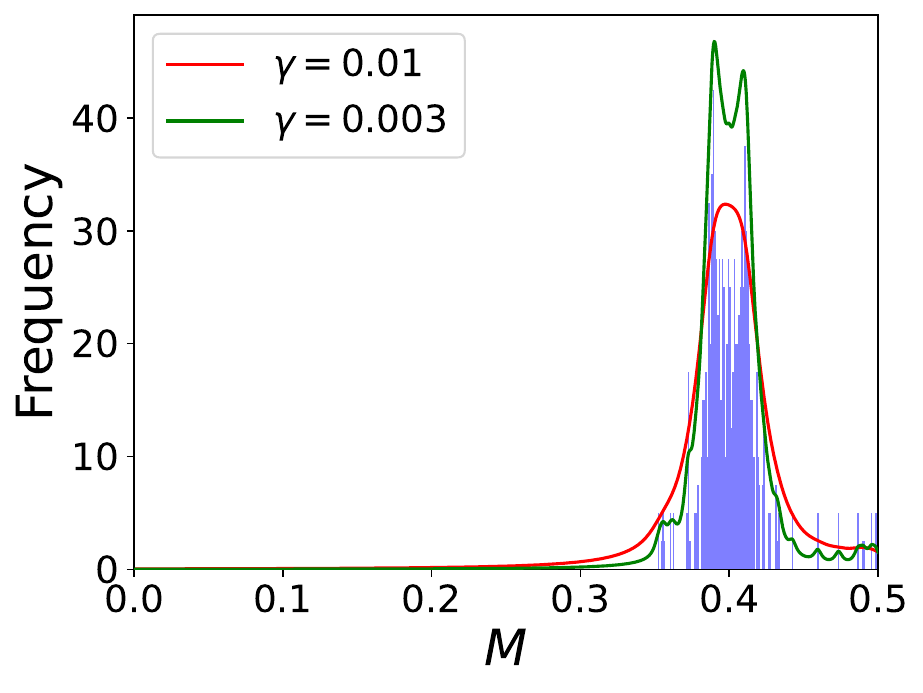}
        \caption{$\lambda=0.8$}
        \label{defectnmr1}
    \end{subfigure}
    \begin{subfigure}{0.23\textwidth}
        \includegraphics[width=\linewidth]{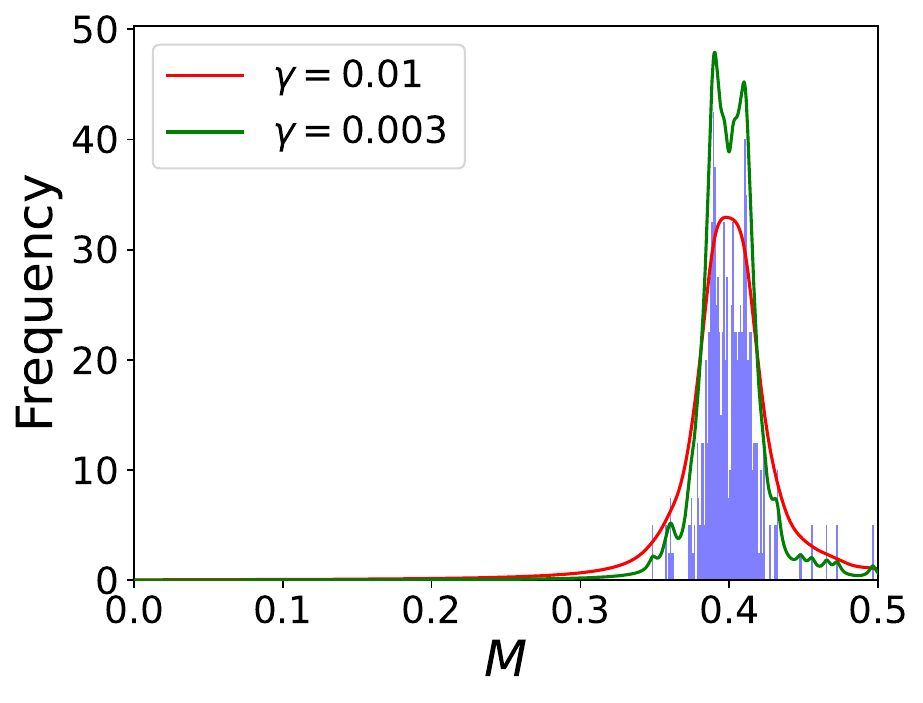}
        \caption{$\lambda=1.0$}
        \label{defectnmr2}
    \end{subfigure}

    \caption{The histogram of magnetization in the spin nematic phase with $J_1/J_2=-0.5$ and $M=0.4$ in the defect model. }
    \label{defectnmrall}
\end{figure}

Figure \ref{defectnmrall} shows the histogram of magnetization in the spin nematic phase corresponding to Fig.\ref{defectmagenall}. 
There is no clear difference in the histograms for $\lambda=1.0$ and $\lambda=0.8$, 
and additional weights around $M\simeq0.5>M_{\rm bulk}=0.4$ are seen for both cases.
These behaviors are different from the experimental observations in the previous NMR study,
where addtional contributions were found for the opposite side of the cental bulk peak, $M<M_{\rm bulk}$.
The inconsistency between the present calculations and the experiment may imply that 
the defect model does not correctly describe the sample used in the experiment 
and the experimentally observed impurity effects have some other origins.
Another possibility is that the exchange interaction is locally 
strengthened in the sample corresponding to $\lambda>1$
in the defect model.
Our poorly converged results for $\lambda>1$ in the defect model are qualitatively similar to Fig.~\ref{nmr3}. 
If there were a certain amount of impurites in a sample, the corresponding histogram would have a tail
in the region $M<M_{\rm sat}$, because of the peaks at small $M$ induced by the impurities.
Although this may be a reasonable possibility, we cannot draw a definite conclusion from the present calculations
and further investigations will be necessary.

\section{Impurity effects in XXZ model}
\label{app:XXZ}

Here, we discuss impurity effects in the XXZ model for a comparison.
We consider the $S=1/2$ XXZ model with an impurity bond described by the Hamiltonian
${\mathcal H}=\mathcal H_{\rm bulk}+\mathcal H_{\rm imp}$, 
\begin{align}
	\mathcal{H}_{\text{bulk}} &=  \sum_{j} J\left( S_j^xS_{j+1}^x+S_j^yS_{j+1}^y\right) + J_zS_j^zS_{j+1}^z  
	- h_z \sum_{j} S^{z}_j,\\
	\mathcal{H}_{\text{imp}} &=  (\lambda-1)\left(J\left( S_{-\frac{1}{2}}^xS_{\frac{1}{2}}^x
	+S_{-\frac{1}{2}}^yS_{\frac{1}{2}}^y\right) + J_zS_{-\frac{1}{2}}^zS_{\frac{1}{2}}^z\right).
\end{align}

\begin{figure}[htb]

    \begin{subfigure}{0.23\textwidth}
        \includegraphics[width=\linewidth]{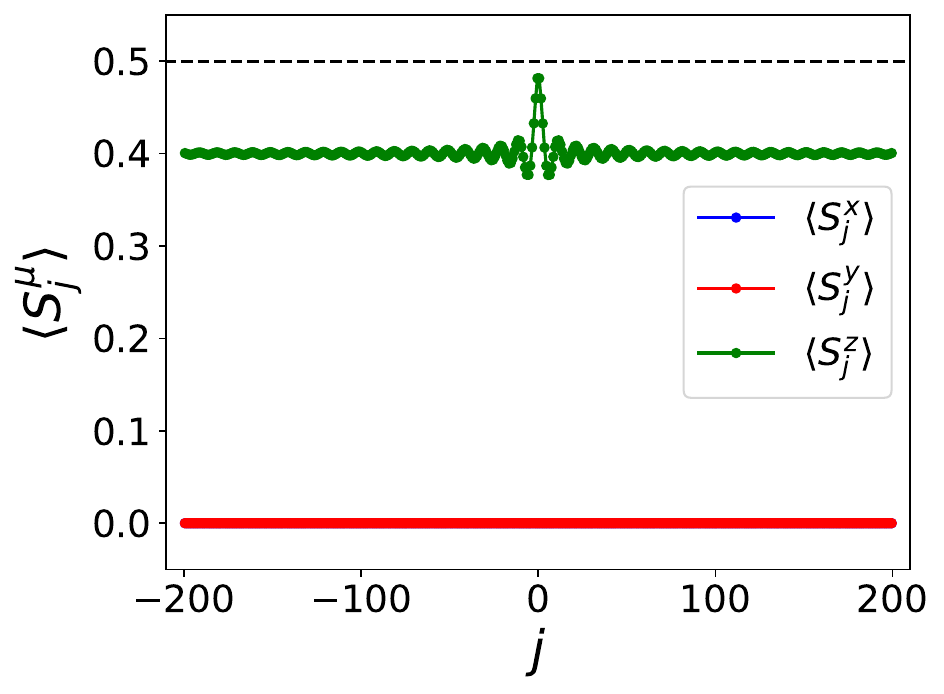}
        \caption{$\lambda=0.8$}
        \label{Heisenmag1}
    \end{subfigure}
    \begin{subfigure}{0.23\textwidth}
        \includegraphics[width=\linewidth]{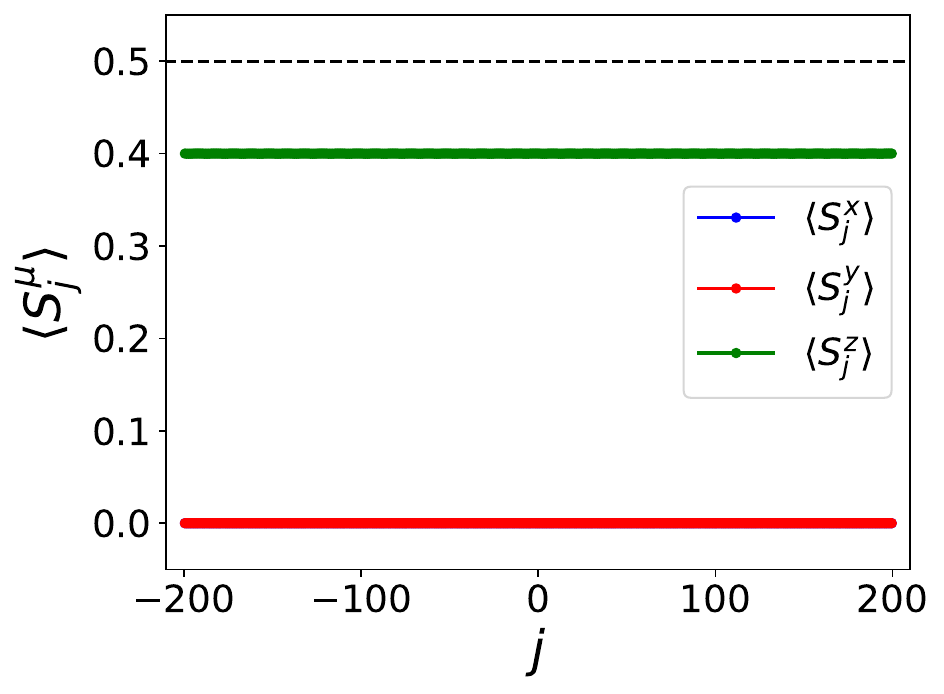}
        \caption{$\lambda=1.0$}
        \label{Heisenmag2}
    \end{subfigure}
    \begin{subfigure}{0.23\textwidth}
        \includegraphics[width=\linewidth]{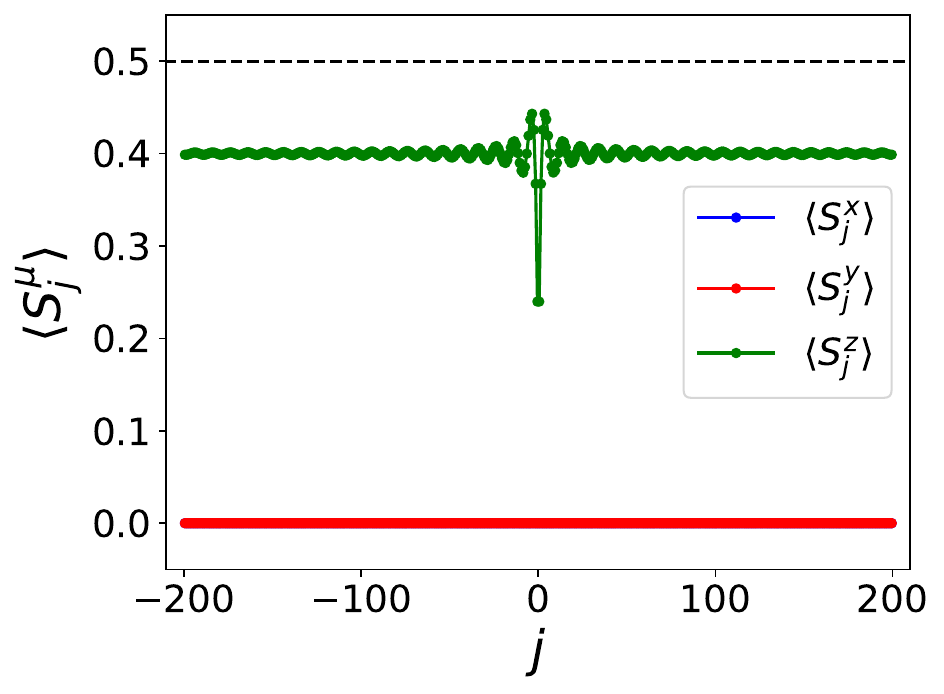}
        \caption{$\lambda=1.2$}
        \label{Heisenmag3}
    \end{subfigure}
    \begin{subfigure}{0.23\textwidth}
        \includegraphics[width=\linewidth]{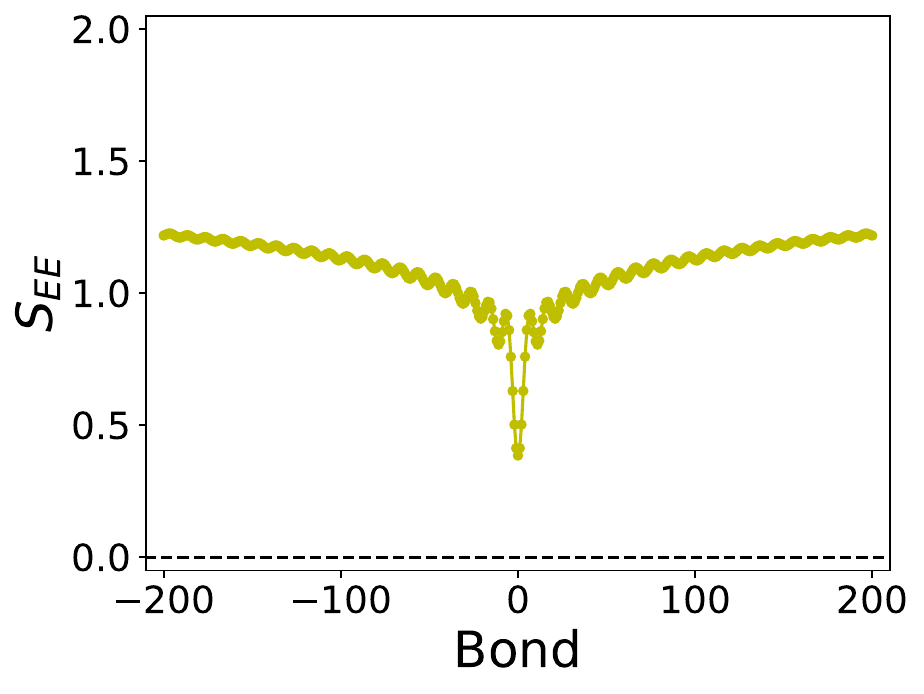}
        \caption{$\lambda=0.8$}
        \label{Heisenen1}
    \end{subfigure}
    \begin{subfigure}{0.23\textwidth}
        \includegraphics[width=\linewidth]{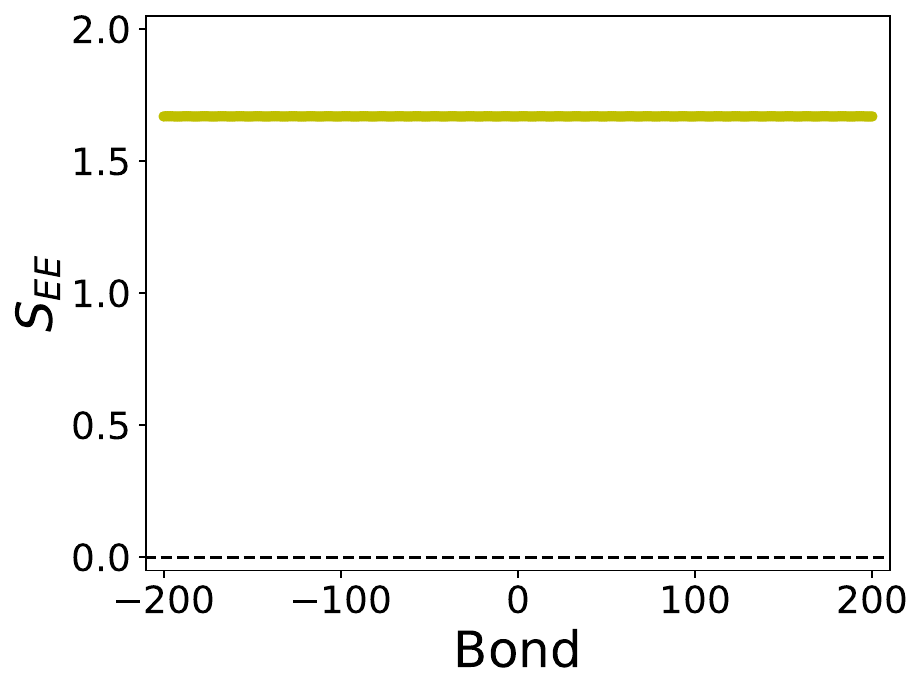}
        \caption{$\lambda=1.0$}
        \label{Heisenen2}
    \end{subfigure}
    \begin{subfigure}{0.23\textwidth}
        \includegraphics[width=\linewidth]{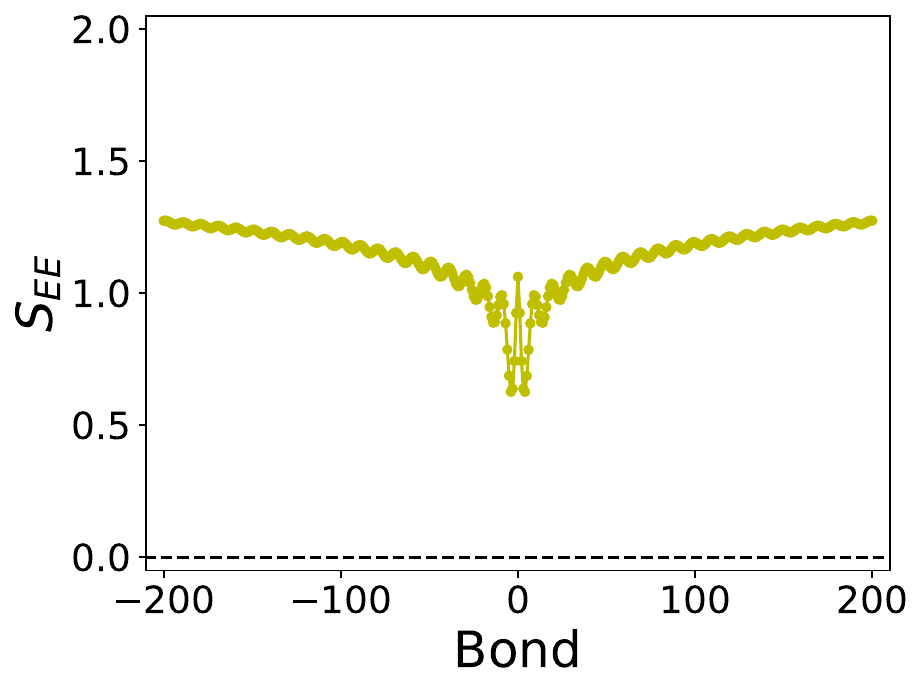}
        \caption{$\lambda=1.2$}
        \label{Heisenen3}
    \end{subfigure}
    
    \caption{(a) (b) (c) The local spin polarization $\braket{S^\mu_j} (\mu = x, y, z)$ and (d) (e) (f) the entanglement entropy $S_{\text{EE}}$ for $L = 400$ with $J_z/J=1$ and $M=0.4$. 
The dashed lines indicate $M = 0.5$ in (a), (b), (c), and $S_{\text{EE}} = 0.0$ in (d), (e), (f), respectively. 
Results for $\lambda=0.8, 1.0, 1.2$ are shown. }
    \label{Heisenmagenall}
\end{figure}

\begin{figure}[htb]
    
    \begin{subfigure}{0.23\textwidth}
        \includegraphics[width=\linewidth]{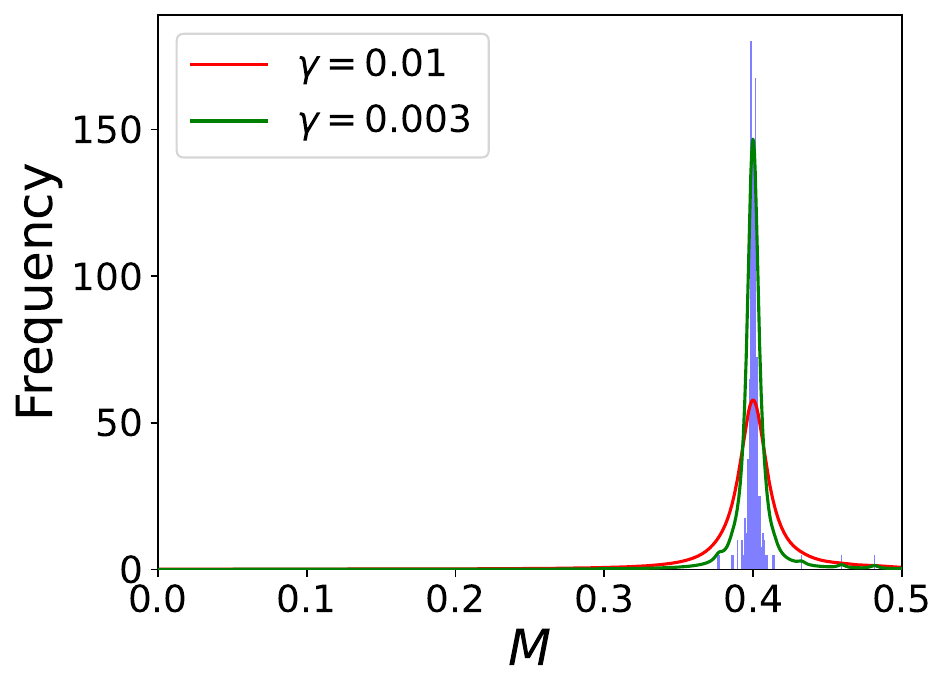}
        \caption{$\lambda=0.8$}
        \label{Heisennmr1}
    \end{subfigure}
    \begin{subfigure}{0.23\textwidth}
        \includegraphics[width=\linewidth]{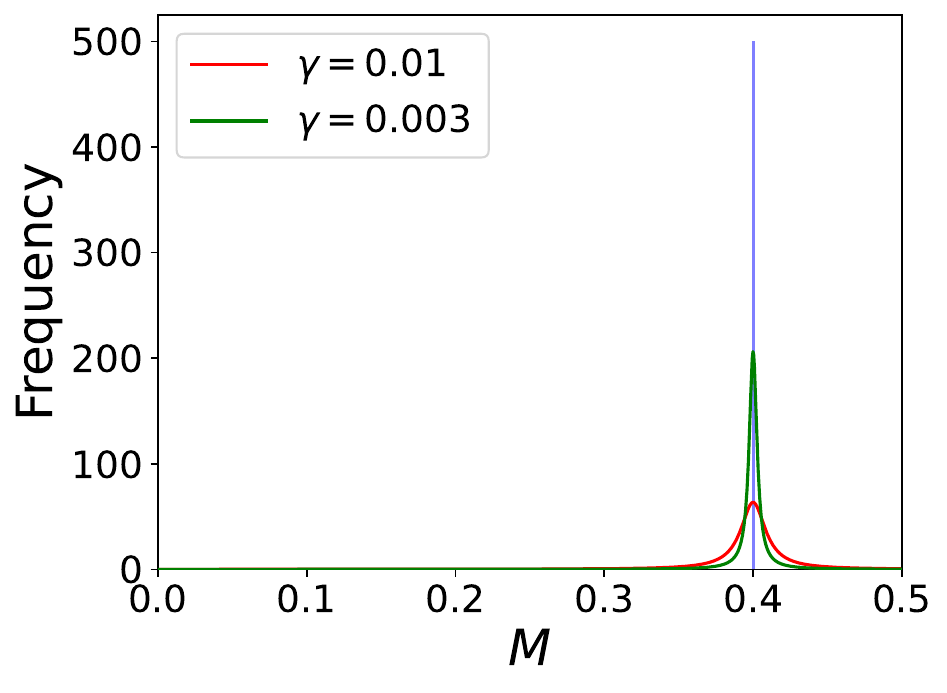}
        \caption{$\lambda=1.0$}
        \label{Heisennmr2}
    \end{subfigure}
    \begin{subfigure}{0.23\textwidth}
        \includegraphics[width=\linewidth]{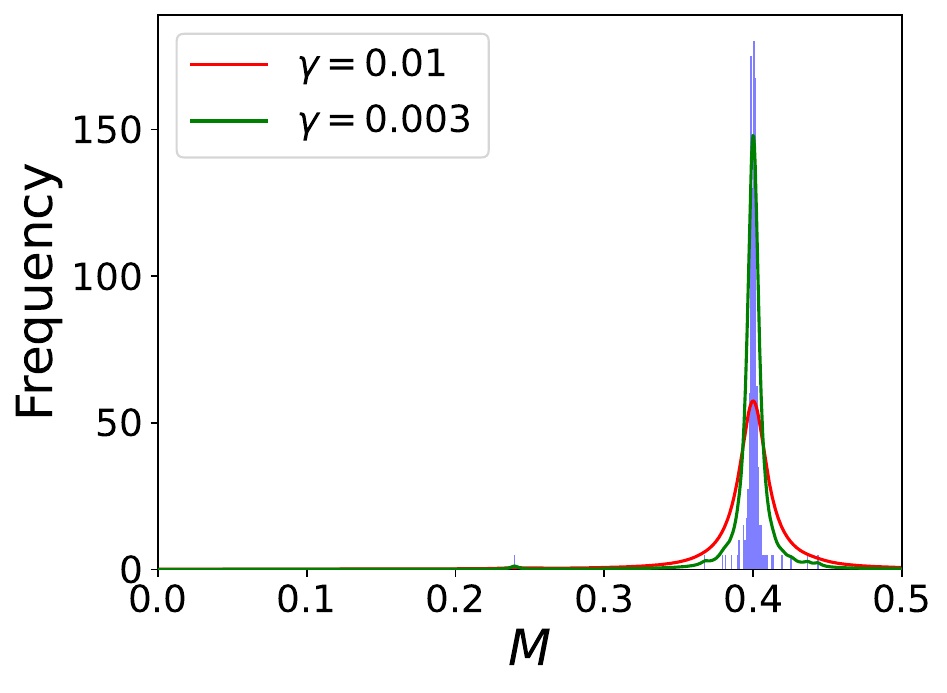}
        \caption{$\lambda=1.2$}
        \label{Heisennmr3}
    \end{subfigure}
    \begin{subfigure}{0.23\textwidth}
        \includegraphics[width=\linewidth]{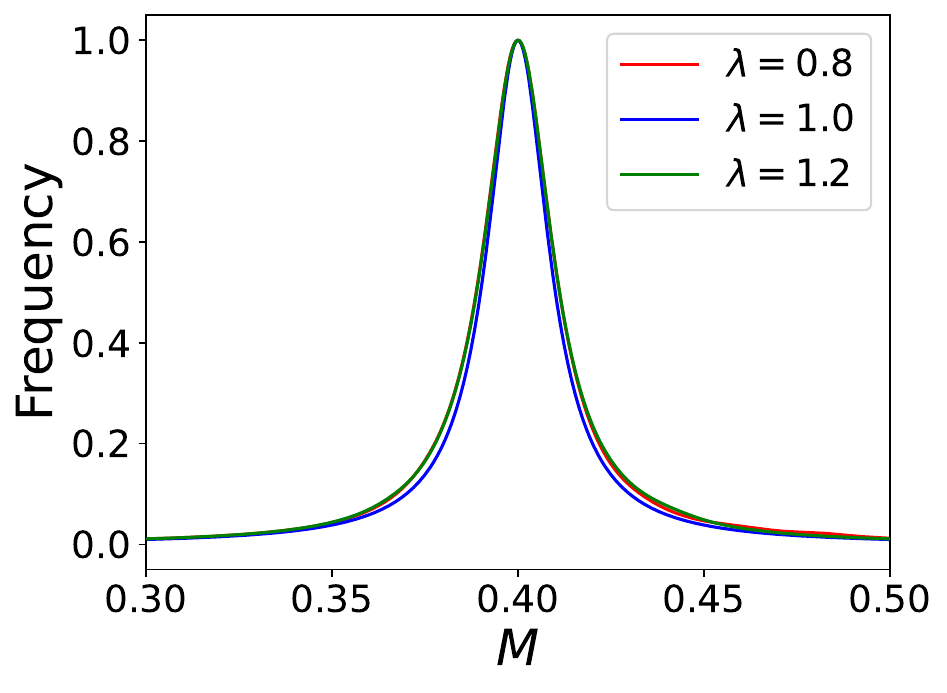}
        \caption{$\gamma=0.01$}
        \label{Heisennmrshape}
    \end{subfigure}

    \caption{(a) (b) (c) Histograms of magnetization of the Heisenberg model with with $J_z/J=1$ and $M=0.4$
	for the $L=400$ window region. The curves are Lorentzian fitting with the phenomenological broadening 
parameter $\gamma$. (d) Normalized fitting curves with $\gamma=0.01$.}
    \label{Heisennmrall}
\end{figure}

Figure \ref{Heisenmagenall}  shows the local magnetization and the entanglement entropy for
$J_z/J=1$ and $M=0.4$.
The magnetization profiles and the entanglement entropy are qualitatively similar to those of the spin nematic state
in the $J_1$-$J_2$ model.
However, there are some quantitative differences.
One point is that the oscillation period is $q=0.2\pi$ which is twice of that for the spin nematic state.
Another point is that line shapes of the histograms are almost unchanged by the impurity,
basically because the TLL parameter is now $K\simeq 0.96$ (estimated from bulk correlation functions) and
the Friedel oscillation in the XXZ model decays faster than in the spin nematic state ($K\simeq 0.61$)
at the same $M=0.4$.
Also, additional peaks are not clearly seen as shown in Fig. \ref{Heisennmrall}.
We perform similar calculations for a smaller magnetization
$M=0.2$ corresponding to the SDW phase of the $J_1$-$J_2$ model.
In this case, we find that local magnetization strongly oscillates around the impurity simiarly to that of the SDW 
state (not shown).

\begin{figure}[htb]
    \begin{subfigure}{0.23\textwidth}
        \includegraphics[width=\linewidth]{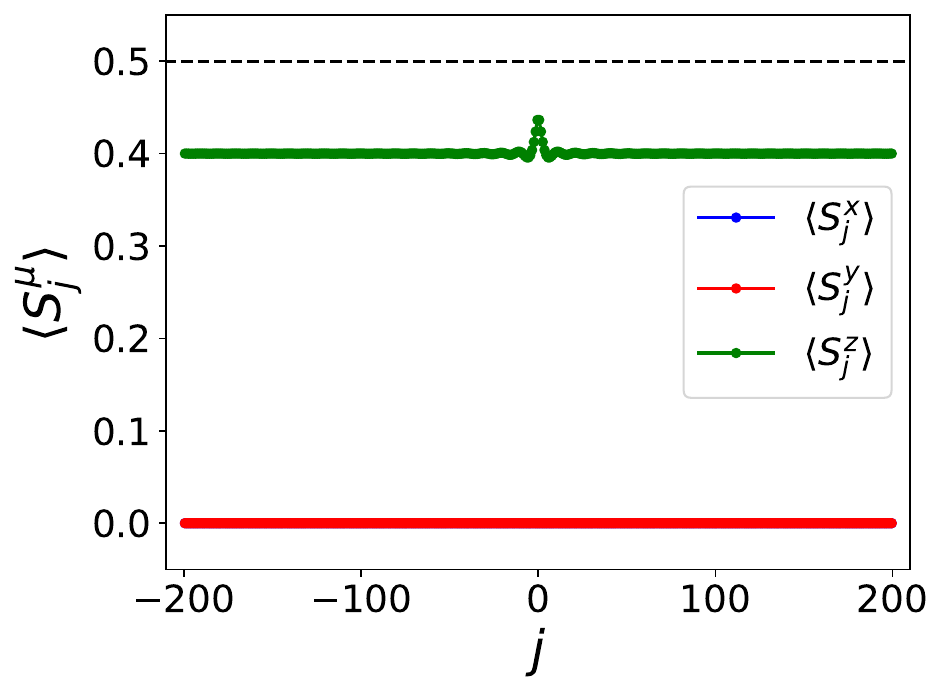}
        \caption{$\lambda=0.8$}
        \label{XXZmag1}
    \end{subfigure}
    \begin{subfigure}{0.23\textwidth}
        \includegraphics[width=\linewidth]{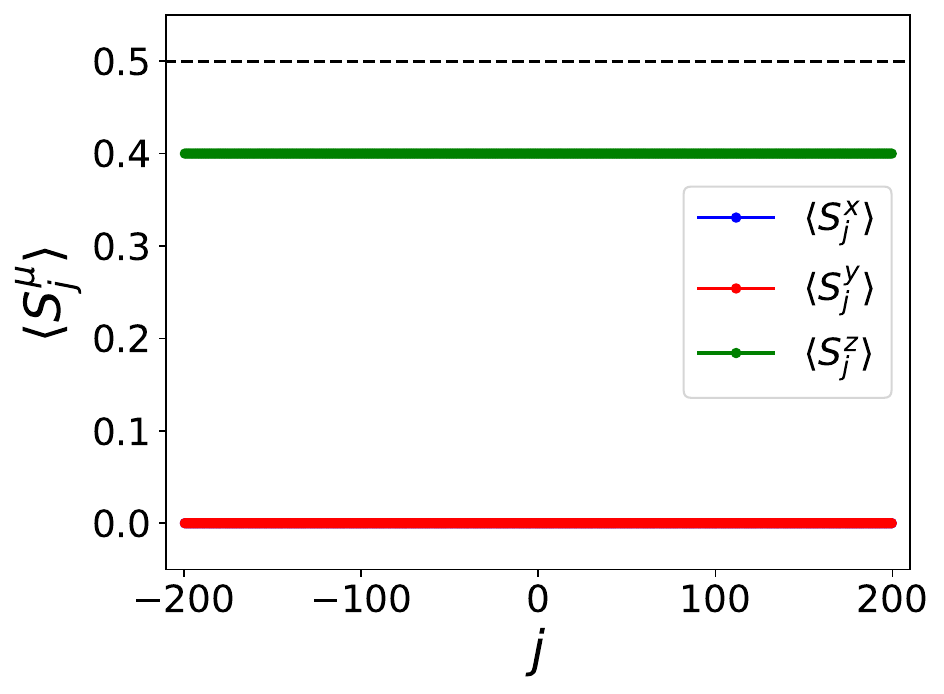}
        \caption{$\lambda=1.0$}
        \label{XXZmag2}
    \end{subfigure}
    \begin{subfigure}{0.23\textwidth}
        \includegraphics[width=\linewidth]{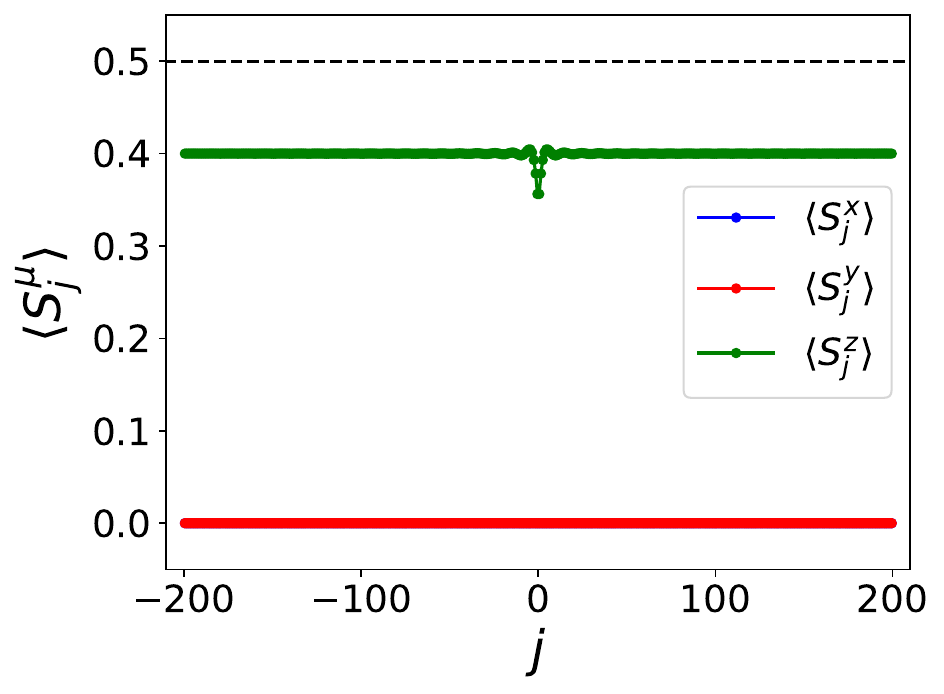}
        \caption{$\lambda=1.2$}
        \label{XXZmag3}
    \end{subfigure}
    \begin{subfigure}{0.23\textwidth}
        \includegraphics[width=\linewidth]{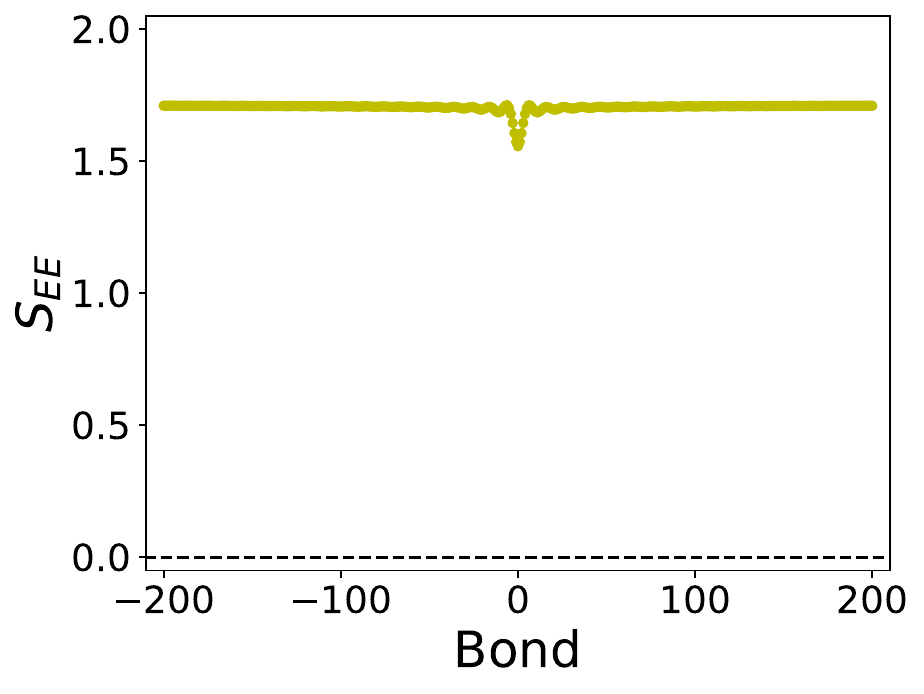}
        \caption{$\lambda=0.8$}
        \label{XXZen1}
    \end{subfigure}
    \begin{subfigure}{0.23\textwidth}
        \includegraphics[width=\linewidth]{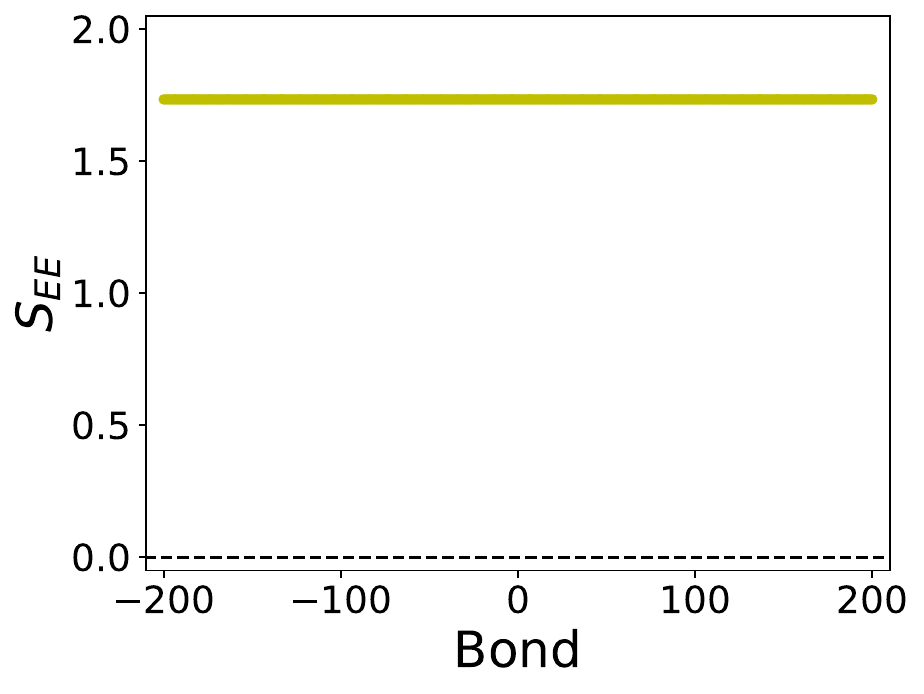}
        \caption{$\lambda=1.0$}
        \label{XXZen2}
    \end{subfigure}
    \begin{subfigure}{0.23\textwidth}
        \includegraphics[width=\linewidth]{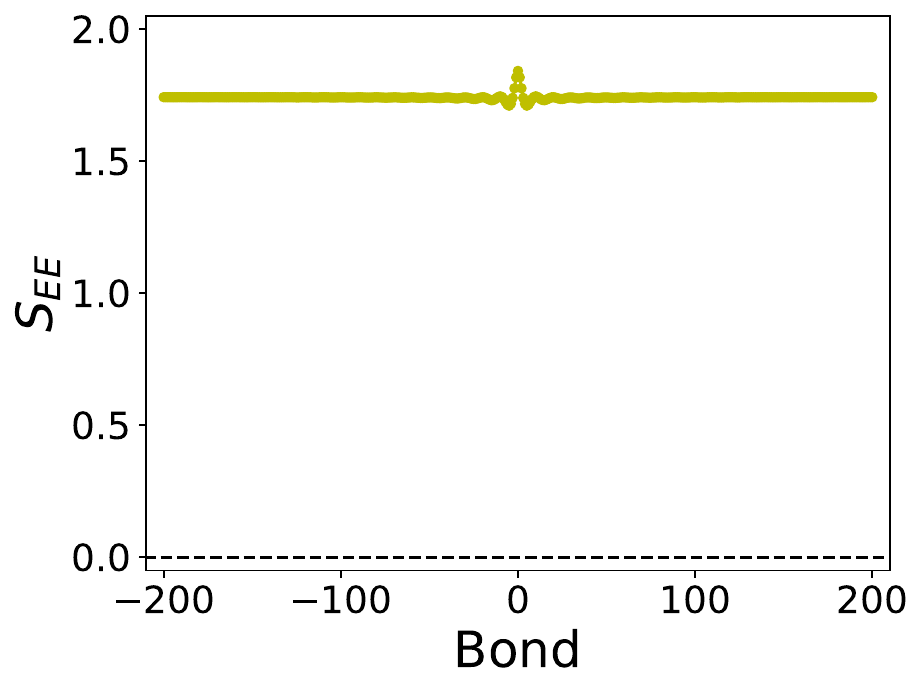}
        \caption{$\lambda=1.2$}
        \label{XXZen3}
    \end{subfigure}
    
    \caption{(a) (b) (c) The local spin polarization $\braket{S^\mu_j} (\mu = x, y, z)$ and (d) (e) (f) the entanglement entropy $S_{\text{EE}}$ for $L = 400$ with $J_z/J=-0.5$ and $M=0.4$. 
The dashed lines indicate $M = 0.5$ in (a), (b), (c), and $S_{\text{EE}} = 0.0$ in (d), (e), (f), respectively. 
Results for $\lambda=0.8, 1.0, 1.2$ are shown. }
    \label{XXZmagenall}
\end{figure}

\begin{figure}[htb]
    
    \begin{subfigure}{0.23\textwidth}
        \includegraphics[width=\linewidth]{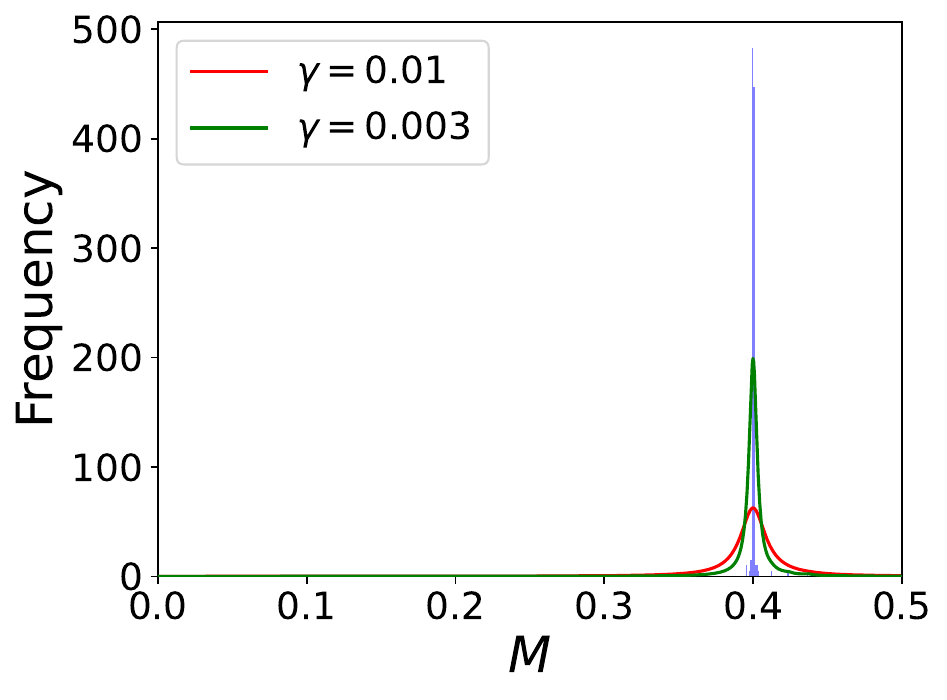}
        \caption{$\lambda=0.8$}
        \label{XXZnmr1}
    \end{subfigure}
    \begin{subfigure}{0.23\textwidth}
        \includegraphics[width=\linewidth]{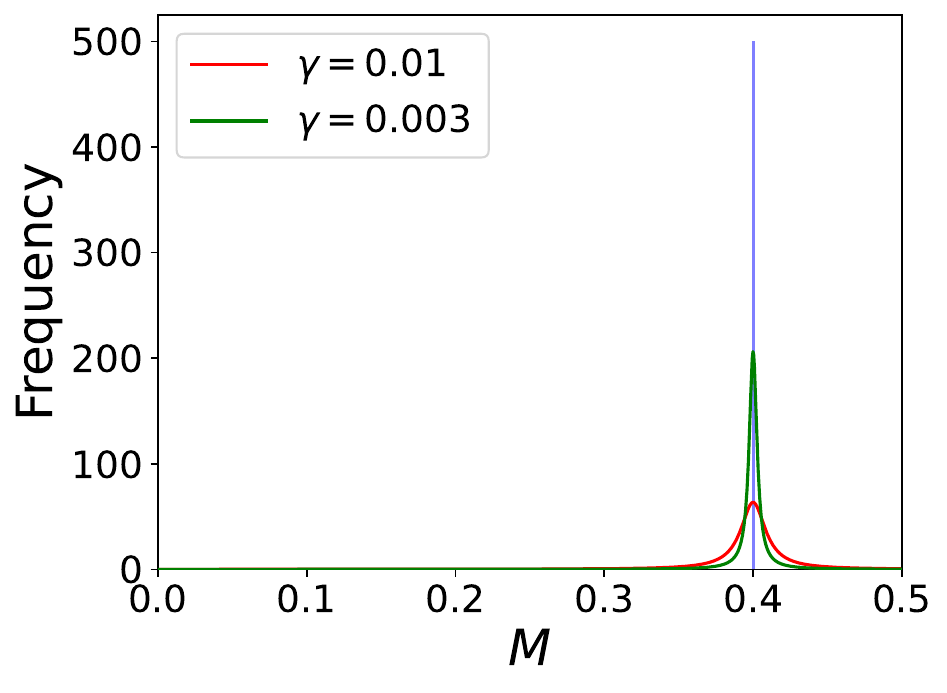}
        \caption{$\lambda=1.0$}
        \label{XXZnmr2}
    \end{subfigure}
    \begin{subfigure}{0.23\textwidth}
        \includegraphics[width=\linewidth]{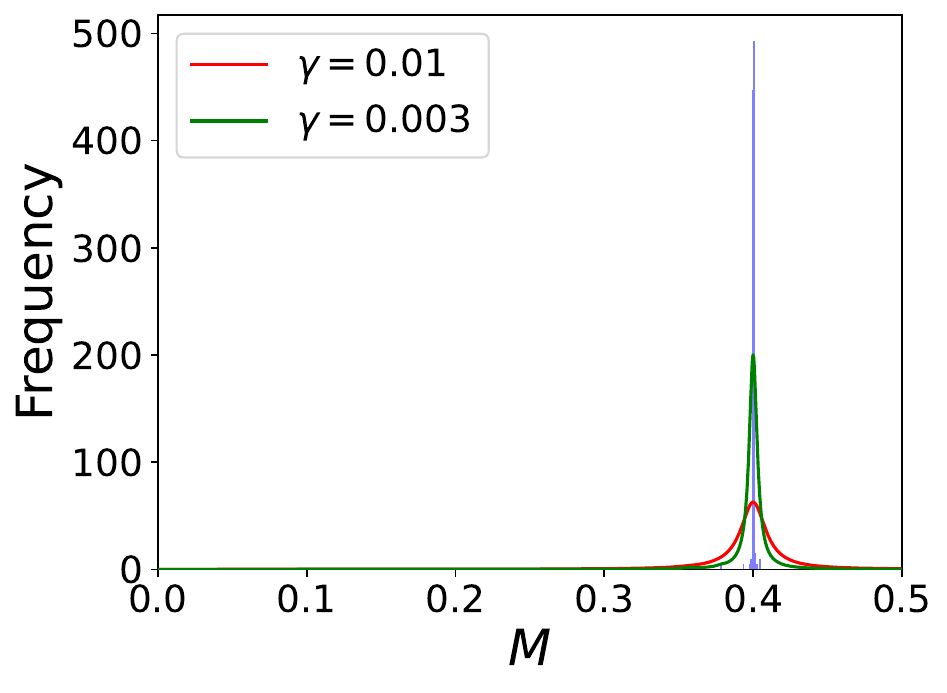}
        \caption{$\lambda=1.2$}
        \label{XXZnmr3}
    \end{subfigure}
    \begin{subfigure}{0.23\textwidth}
        \includegraphics[width=\linewidth]{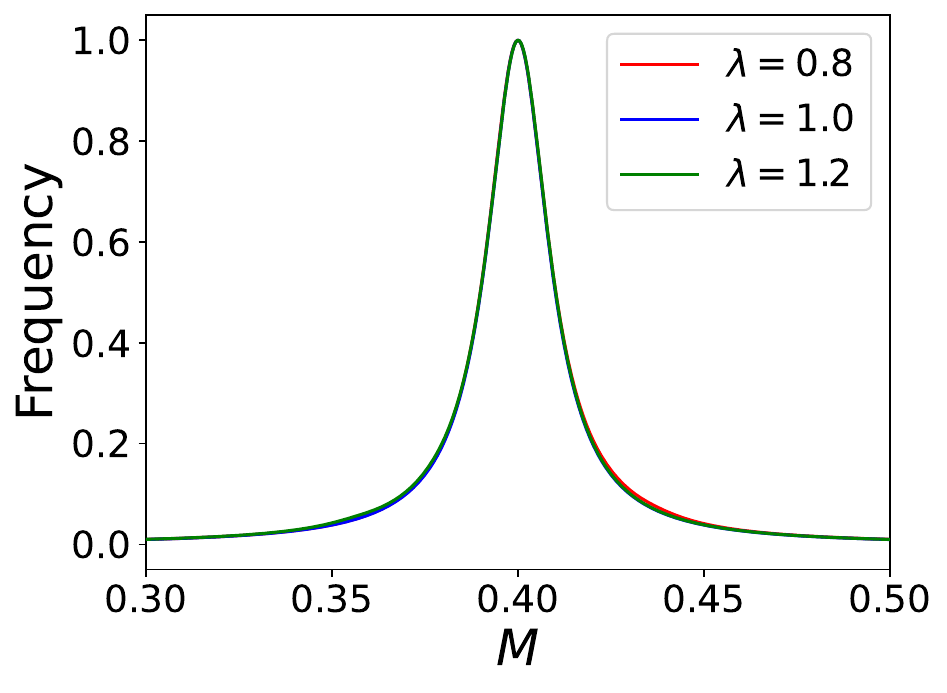}
        \caption{$\gamma=0.01$}
        \label{XXZnmrshape}
    \end{subfigure}

    \caption{(a) (b) (c) Histograms of magnetization of the XXZ model with with $J_z/J=-0.5$ and $M=0.4$
	for the $L=400$ window region. The curves are Lorentzian fitting with the phenomenological broadening 
parameter $\gamma$. (d) Normalized fitting curves with $\gamma=0.01$.}
    \label{XXZnmrall}
\end{figure}

We also present results for $J_z/J=-0.5$ and $M=0.4$ in Fig. \ref{XXZmagenall}. 
In this case, the impurity effects are strongly suppressed in accordance with the bosonization result that
the perturbation potential is irrelevant and renormalized to zero.
As a result, the histograms remain very sharp as shown in Fig.~\ref{XXZnmrall}.




\section{Lorentzian fitting}
\label{app:Lorentzian}
The Lorentzian fitting of the magnetization histogram $H(M)$ is performed as follows.
First, we discretize the range of the magnetization $[0,0.5]$ into $N=500$ intervals $\{[M_{k-1},M_{k}]\}_{k=1}^N$ 
with $M_k=0.5k/N$, and make a histogram $H(M)$.
We normalize the histogram so that $(1/N)\sum_{j=1}^{N}H(M_j)=1$.
The histogram is reprensented by $H(M)=\sum_{k=1}^N c_k\delta_{\varepsilon}(M - \tilde{M}_k)$ as a weighted sum
of the approximate delta functions 
\begin{align}
\delta_{\varepsilon}(x)=\left\{
\begin{array}{ll}
0 & (|x|>\varepsilon/2) \\
1/\varepsilon & (|x|\leq \varepsilon/2)
\end{array}\right.
,
\end{align}
where $\varepsilon=0.5/N=0.001$ and $\tilde{M}=\left( M_{k-1} + M_k \right)/2$.
Then, we replace the approximate delta function with the Lorentzian function, 
\begin{align}
\delta_{\varepsilon}(M - \tilde{M}_k)\to \Delta_{\gamma}(M - \tilde{M}_k)=\frac{1}{\pi}\frac{\gamma}{(M - \tilde{M}_k)^2+\gamma^2},
\end{align}
where $\gamma$ is a phenomenological broadening parameter.
The total fitting function is given by
$F(M)=\sum_{k=1}^N c_k \Delta_{\gamma}(M - \tilde{M}_k)$.

\clearpage
\bibliography{Ref}
\bibliographystyle{apsrev4-1}

\end{document}